\documentclass[tightenlines,nofootinbib,floatfix,notitlepage,superscriptaddress,11pt]{revtex4-1}

\usepackage{tcolorbox}
\tcbuselibrary{theorems}
\usepackage{epsfig}
\usepackage{graphicx}
\usepackage{multirow}
\usepackage[inline]{enumitem}
\usepackage{color}
\usepackage[dvipsnames]{xcolor}
\usepackage{amsmath,amssymb}
\usepackage{tabularx}
\usepackage[pdfencoding=auto, psdextra]{hyperref}
\usepackage{units}
\usepackage{soul}
\usepackage[T1]{fontenc}
\usepackage{ulem} 
\usepackage{aas_macros}
\usepackage{upgreek}
\usepackage{mathrsfs}
\usepackage{bbm}
\usepackage{bm}
\usepackage{bbold} 
\usepackage{enumitem}
\usepackage{comment}
\usepackage{slashed}
\usepackage[capitalise]{cleveref}
\usepackage{xspace}

\newcommand{\be}{\begin{equation}}
\newcommand{\ee}{\end{equation}}
\newcommand{\bea}{\begin{eqnarray}}
\newcommand{\eea}{\end{eqnarray}}

\newcommand{\vrel}{v_{\rm rel}}
\newcommand{\nuloss}{\nu_{\rm loss}}

\newcommand{\GeV}{\,\mathrm{GeV}}

\newcommand{\cmg}{\; \textrm{cm}^2/\textrm{g}}

\newcommand{\kms}{\, \mathrm{km/s}}

\newcommand{\Gammacool}{\Gamma_{\rm cool}}
\newcommand{\gravothermalsidm}{\texttt{GravothermalSIDM}\xspace}
\newcommand{\spherical}{\texttt{SPHerical}\xspace}

\AddToHook{cmd/appendix/before}{\crefalias{section}{appendix}}

\hypersetup{
  colorlinks   = true, 
  urlcolor     = Emerald, 
  linkcolor    = Emerald, 
  citecolor   = Emerald 
}

\begin{document}

\title{Core-collapse universality for inelastic self-interacting dark matter halos}

\author{Evan Batteas}
\email{batteas@wustl.edu}
\affiliation{Department of Physics, Washington University in St. Louis,
St. Louis, MO 63130, U.S.A.}
\author{Alexander Brisebois}
\email{alex.brisebois@mail.utoronto.ca}
\affiliation{David A. Dunlap Department of Astronomy and Astrophysics, University of Toronto, Ontario, M5S 3H4, Canada}
\author{Maheen Hemani}
\email{mh12@yorku.ca}
\affiliation{Department of Physics and Astronomy, York University, Toronto, Ontario, M3J 1P3, Canada}
\author{Maximilian Heyne}
\affiliation{Institute for Theoretical Physics, Goethe University, 60438 Frankfurt am Main, Germany}
\author{\\ J. Leo Kim}
\email{jlkim@yorku.ca}
\affiliation{Department of Physics and Astronomy, York University, Toronto, Ontario, M3J 1P3, Canada}
\author{Lukas Lehmann}
\affiliation{Institute for Theoretical Physics, Goethe University, 60438 Frankfurt am Main, Germany}
\author{Laura Sagunski}
\affiliation{Institute for Theoretical Physics, Goethe University, 60438 Frankfurt am Main, Germany}
\author{Aster Schnell}
\affiliation{Department of Physics and Astronomy, York University, Toronto, Ontario, M3J 1P3, Canada}
\author{Sean Tulin}
\email{stulin@yorku.ca}
\affiliation{Department of Physics and Astronomy, York University, Toronto, Ontario, M3J 1P3, Canada}

\date{\today}

\begin{abstract}
We study the gravothermal collapse of inelastic dark matter halos with velocity-dependent elastic and dissipative self-interactions.
Within the gravothermal fluid framework, we derive the effective cooling rate for dark matter upscattering to an excited state followed by decay back to the ground state, with emitted states carrying off energy from the halo.
We solve the gravothermal fluid equations numerically to simulate spherically symmetric halos over the core-growing and collapse phases of their evolution.
While halos with purely elastic self-interactions have universal evolution trajectories, inelastic dark matter halos exhibit much more diversity in their evolution.
Nevertheless, inelastic halo dynamics contain a hidden parametric universality from which we derive a semi-analytic formula for calculating the halo collapse time for arbitrary velocity-dependent elastic and inelastic cross sections without the need for simulations.
\end{abstract}

\maketitle



\section{Introduction \label{sec:intro}} 

The fundamental nature of dark matter (DM) is unknown despite decades of observational evidence that DM comprises approximately 85\% of the matter content of the universe \cite{Planck:2018vyg}.
The standard paradigm of collisionless cold dark matter (CDM) has found great success in matching our observations on galactic and cosmological scales, though challenges remain on smaller  scales~\cite{Weinberg:2013aya,Bullock:2017xww}.
Many works have suggested that the proper modeling of baryonic physics within CDM simulations can alleviate these tensions~\cite{Navarro:1996bv,Governato:2012fa,Pontzen:2011ty,Zolotov:2012xd,2018MNRAS.473.4392S,Cruz:2025}. 
Nevertheless, models beyond the CDM paradigm have gained traction not only to resolve these small-scale issues, but also because non-minimal dark sectors lead to interesting phenomena that can be confronted against observations.

Self-interacting dark matter (SIDM) is a class of DM models in which the collisionless assumption of CDM is relaxed due to the presence of dark interactions~(see Refs.~\cite{Tulin:2017ara,Adhikari:2022sbh} for reviews).
In its original form~\cite{Spergel:1999mh}, SIDM halos form constant-density cores (in contrast to cuspy CDM halos) due to elastic scattering between DM particles that transfers heat inward to the halo center.
The core-cusp~\cite{Moore:1994yx,Flores:1994gz}, too-big-to-fail~\cite{BoylanKolchin:2011de,BoylanKolchin:2011dk}, and diversity~\cite{KuziodeNaray:2009oon,Oman:2015xda} problems can all be elegantly explained if the scattering cross section per unit DM mass is $\sigma/m \sim 0.5 - 50 \; {\rm cm^2/g}$~\cite{Vogelsberger:2012ku,Kaplinghat:2015aga,Elbert:2014bma,Kamada:2016euw,Ren:2018jpt}.
If $\sigma/m \gg 50 \; {\rm cm^2/g}$, heat flow in SIDM halos reverses and gravothermal core-collapse may ensue over the age of the Universe~\cite{Colin:2002nk}.

Beyond minimal SIDM with purely elastic scattering, dissipative self-interacting dark matter (dSIDM) extends the collisional phenomenology of DM by introducing inelastic scattering. 
For example, with an interaction rate far larger than weak-scale physics typically associated with CDM, many SIDM models invoke light mediator exchange from hidden sector dynamics~\cite{Buckley:2009in,Feng:2009mn,Feng:2009hw,Bringmann:2013vra,Tulin:2013teo,Cline:2013pca,Boddy:2014yra,Boddy:2016bbu}.
Collisions may also produce on-shell light mediators that escape and carry away energy from the halo.

Dissipative processes allow DM halos to radiate away kinetic energy rather than simply redistributing it (as for elastic scattering), resulting in distinct phenomenology from both CDM and elastic-SIDM.
This includes accelerated core-collapse~\cite{Essig:2018pzq, Huo:2019yhk}, forming dark disks and other modified halo structure~\cite{Fan:2013tia, Fan:2013yva, Roy:2023zar, Roy:2024bcu, Gemmell:2023trd, Shen:2021frv, Shen:2022opd, Low:2025llz}, 
forming dark compact objects~\cite{Buckley:2017ttd, Chang:2018bgx, Gurian:2022nbx, Bramante:2023ddr, Bramante:2024pyc,  Litterer:2025quq}, and seeding supermassive black holes at high redshifts~\cite{DAmico:2017lqj, Xiao_2021_SMBHSeeds, Buckley:2024eoe, shen2025massiveblackholesseeded}. 
Models for dSIDM proposed in the literature include atomic dark matter~\cite{Kaplan:2009de, CyrRacine:2012fz}, mirror dark matter~\cite{Berezhiani:1995am}, and dark plasma models~\cite{Clarke_2016}. 
Furthermore, the LUX-ZEPLIN (LZ) experiment's recent detection of a single, large-recoil event consistent with nuclear scattering~\cite{LZ:2026axp} has put inelastic models of DM in the limelight (e.g.~\cite{Asadi:2026iot, Pospelov:2026ewn, Rodd:2026tyn, Yamashita:2026ump, Dent:2026bji}).

In this work, we provide a framework for exploring the astrophysical phenomenology of realistic particle models with velocity-dependent elastic and inelastic scattering.
$N$-body simulations remain the state of the art for modeling nonlinear structure dynamics and several recent studies have incorporated inelastic or dissipative interactions for DM~\cite{Huo:2019yhk, Shen:2021frv, Shen:2022opd, Low:2025llz,Roy:2023zar, Gemmell:2023trd, Roy:2024bcu, ONeil:2022szc, Vogelsberger:2018bok}. 
However, $N$-body simulations are also computationally expensive, which makes it challenging to explore the landscape of particle parameters and different astrophysical systems.

Instead, we perform simulations using the gravothermal fluid framework for SIDM halos.
In this setup, DM is modeled as a self-gravitating fluid, with elastic self-interactions encoded through heat conductivity~\cite{Gnedin:2000ea,Balberg:2002ue,Ahn:2004xt,Koda:2011yb,Nishikawa:2019lsc} and including velocity-dependent scattering~\cite{Outmezguine:2022bhq,Gad-Nasr:2023gvf}.
Dissipative scattering is treated as an effective cooling term in the gravothermal framework~\cite{Essig:2018pzq}.
We utilize two simulation packages, the \gravothermalsidm~\cite{Nishikawa:2019lsc, Outmezguine:2022bhq, Gad-Nasr:2023gvf} and \spherical~\cite{blaff_spherical,SPHerical_paper} codes, to solve the gravothermal fluid equations with heat conduction and cooling, suitably generalized to include velocity-dependent elastic and inelastic scattering.

We consider a minimal model for inelastic DM that is well-known in the literature~\cite{Tucker-Smith:2001myb,Tucker-Smith:2004mxa,Finkbeiner:2007kk, Cui:2009xq, Schutz:2014nka, Barello:2014uda, Bramante:2016rdh, Alvarez:2019nwt}.
We have a Dirac fermion $\Psi$ and a light $U(1)$ gauge boson, the hidden sector analog of electrodynamics, with gauge coupling $g_\chi$.
Spontaneous symmetry breaking splits the Dirac fermion into two Majorana fermions, a ground state $\chi_{1}$ and an excited state $\chi_2$, with a naturally small mass splitting $\delta = m_2 - m_1 \ll m_1$, and also generates a mediator mass $m_\phi$.
The interaction Lagrangian becomes purely off-diagonal
\begin{align}
    \mathcal{L}_\mathrm{int} \supset \frac{i}{2} g_\chi \bar{\chi}_2 \gamma^\mu \chi_1 \phi_\mu + \text{h.c.} \label{eq:int_Lagrangian} \, .
\end{align}
We take only $\chi_1$ to be cosmologically stable, while $\chi_2$ is presumed short-lived and rapidly decaying.\footnote{If $\delta > m_\phi$, then $\chi_2 \to \chi_1 \phi$ can occur rapidly. We assume $\phi$ particles are not reabsorbed elsewhere in the halo, i.e., no dark radiative transfer.
If $\delta < m_\phi$, decays can still proceed through off-shell $\phi$ if we introduce sufficiently light hidden sector states. 
We remain agnostic about such states, provided they freely escape the halo.
\label{fn:decay}}
Dissipation arises as a two-step process of upscattering $\chi_1 \chi_1 \to \chi_2 \chi_2$ followed by a prompt decay of $\chi_2$ back to $\chi_1$, with $\phi$ (real or off-shell) carrying off energy.
Since Eq.~\eqref{eq:int_Lagrangian} is purely off-diagonal, inelastic scattering arises via tree-level $\phi$ exchange, while elastic scattering $\chi_1 \chi_1 \to \chi_1 \chi_1$ occurs at loop order.
Elastic and inelastic cross sections are calculated self-consistently from the underlying particle parameters using a quantum mechanical two-state partial wave analysis~\cite{Blennow:2016gde,Alvarez:2019nwt}.
These numerical calculations will be discussed further in our companion paper~\cite{quantum_team_paper}.

Our main results demonstrate how the mass splitting $\delta$ can dramatically accelerate or slow the onset of gravothermal collapse in halos compared with purely elastic SIDM with $\delta = 0$.
At first, increasing $\delta$ decreases the halo collapse time as the energy loss per scattering increases, reaching a minimum when $\delta$ is comparable to the typical kinetic energy per DM particle~\cite{Essig:2018pzq}.
Thereafter, larger values of $\delta$ increase the collapse time as upscattering becomes Boltzmann-suppressed. 
Elastic scattering becomes suppressed as well, since the loop process proceeds only through heavy $\chi_2$ states, and the collapse time can be far longer than the $\delta=0$ case.

Even within this minimal two-state dSIDM model, elastic and inelastic scattering cross sections exhibit a wide range of velocity dependencies~\cite{Blennow:2016gde,Alvarez:2019nwt,quantum_team_paper}.
For highly idealized simulations such as ours, it becomes prohibitively expensive to scan over the parameter landscape for a variety of astrophysical systems.
In this work, we derive a universal semi-analytic formula for calculating the collapse time of dSIDM halos for velocity-dependent cross sections for a wide range of model parameters and different halo systems.
Ref.~\cite{Essig:2018pzq} used the absence of collapsed halos in dwarf galaxies to constrain dSIDM with constant cross sections.
Our results allow such constraints to be applied across dSIDM parameter space without running numerical simulations at all.
We will discuss these constraints in future work~\cite{constraints_team_paper}.

The remainder of this paper is structured as follows. 
In \cref{sec:gravothermal}, we carefully derive the gravothermal fluid equations for our dSIDM model.
Starting with the two-fluid Navier-Stokes equations for $\chi_1$ and $\chi_2$, we calculate the interaction terms from upscattering and decay from the Boltzmann collision terms, which we then reduce to an effective one-fluid description in the limit that $\chi_2$ is short-lived.
In \cref{sec:sim1}, we present our numerical simulations of gravothermal dSIDM halos for a number of benchmark models.
Our results focus on halos representative of dwarf galaxies, showing the trajectory of halo central densities through core growth and subsequent collapse.
We also demonstrate the numerical convergence and mutual agreement of our simulation codes.
In \cref{sec:timescales}, we derive and test our semi-analytic calculation for collapse time for dSIDM halos with velocity-dependent cross sections. 
Finally, we conclude in \cref{sec:conclude}. 

\section{Gravothermal equations \label{sec:gravothermal}} 

\subsection{Navier-Stokes equations for a two-component fluid} \label{sec:two-fluid}

The gravothermal fluid equations for the evolution of a single-component dark matter fluid can be modeled by the Navier-Stokes equations for a single (heat) conducting fluid \cite{Pitaevskii1981Physical, Kolb:1990vq}:
\begin{align}
    \label{eq:euler1}
    D_t\rho &= - \rho \boldsymbol{\nabla} \cdot \mathbf{v} \\
    \label{eq:euler2}
    D_t \mathbf v &= -  \frac{\boldsymbol{\nabla} P}{\rho}  +  \mathbf{g} \\
    D_t e &= - \frac{P}{\rho} \boldsymbol{\nabla} \cdot \mathbf{v} + \frac{1}{\rho} \boldsymbol\nabla \cdot (\tilde{\kappa} \boldsymbol{\nabla}e) 
    ,\label{eq:euler3}
\end{align}
where we have the fluid mass density $\rho$, bulk fluid velocity $\mathbf{v}$, internal specific energy $e$, pressure $P$, gravitational acceleration $\mathbf{g} = - \boldsymbol{\nabla}\Phi$, and total gravitational potential $\Phi$.
Internal energy is related to the (1D) velocity dispersion $\nu$ and the temperature $T$ per unit mass $m$ by $e = \frac{3}{2} \nu^2 = \frac{3}{2} T / m$ (we set $k_B=1$). 
Fourier's law of conduction states that the heat flux is $\mathbf{q} = - \kappa \boldsymbol{\nabla} T$, with heat conductivity $\kappa$, which describes energy transfer through elastic self-interactions.
Here we use the rescaled heat conductivity $\tilde\kappa = \frac{2}{3} m \kappa$ to write \cref{eq:euler3} in terms of $e$ rather than $T$.

In this paper, since we have a system of interacting fluids (namely the $\chi_1$ and $\chi_2$ particles), we must generalize the above expressions. This results in two copies of each expression above, one for $\chi_1$ and one for $\chi_2$, along with additional interaction terms between the two species. Therefore one must have
\begin{align}
    \label{eq:euler1_general_i}
    D_t\rho_i &= - \rho_i \boldsymbol{\nabla} \cdot \mathbf{v}_i + \mathcal{S}_{\mathrm{decay},i} + \mathcal{S}_{\mathrm{up},i} \\
    \label{eq:euler2_general_i}
    D_t \mathbf v_i &= -  \frac{\boldsymbol{\nabla} P_i}{\rho_i}  +  \mathbf{g} + \bm{\mathcal{S}}_{\mathrm{decay},i} + \bm{\mathcal{S}}_{\mathrm{up},i} \\
    D_t e_i &= - \frac{P_i}{\rho_i} \boldsymbol{\nabla} \cdot \mathbf{v}_i + \frac{1}{\rho_i} \boldsymbol\nabla \cdot (\tilde{\kappa}_i \boldsymbol{\nabla}e_i) + \tilde{\mathcal{S}}_{\mathrm{decay}, i} + \tilde{\mathcal{S}}_{\mathrm{up}, i}
    ,\label{eq:euler3_general_i}
\end{align}
for each species $i=1,2$. 

Inelastic transitions between species are encoded by source terms $\mathcal{S}_{i}, \bm{\mathcal{S}}_{i}$, and $\tilde{\mathcal{S}}_i$ for the fluid density, velocity, and internal energy respectively. 
These terms follow from the Boltzmann collision term and can be written generically for a species $i$ as 
\begin{align}
    \tilde{\bm{\mathcal{S}}}_{\mathrm{decay}, i} &= \pm \int d\Pi_p d\Pi_q \prod_j d\Pi_{k_j} (2 \pi)^4 \delta^4 \left(q - p - \sum_j k_j \right) \left( \sum |\mathcal{M}|^2 \right) \mathbf{g}(\mathbf{u}_i) f_2 (\mathbf{q}), \label{eq:source_decay} \\
    \tilde{\bm{\mathcal{S}}}_{\mathrm{up}, i} &= \pm \int d\Pi_{p} d\Pi_{p'} d\Pi_q d\Pi_{q'} (2 \pi)^4 \delta^4(p + p' - q - q') \left( \sum |\mathcal{M}|^2 \right) \mathbf{g}(\mathbf{u}_i) f_1(\mathbf{p}) f_1(\mathbf{p}'), \label{eq:source_upscat}
\end{align}
where $d\Pi_q = d^3\mathbf{q}/[(2\pi)^3 2 E_q]$, defined similarly for $\mathbf{p}$ and $\mathbf{k}$, and $f_i$ is the phase space distribution function. 
We denote $\mathbf{q}, \mathbf{p}, \mathbf{k}_j$ (and primed values as well) as the 3-momenta for $\chi_2$, $\chi_1$, and the outgoing decay products, respectively. 
The quantity $\mathbf{g}(\mathbf{u}_i)$ (and therefore $\tilde{\bm{\mathcal{S}}}$) is a scalar or vector function that takes moments of the particle velocity $\mathbf{u}_i$ for species $i$. 

In \cref{eq:source_decay}, we allow for a general decay $\chi_2 \to \chi_1 X$. If on-shell decay $\chi_2 \to \chi_1 \phi$ is allowed, the process is a simple two-body decay. 
However, if $\delta < m_\phi$ (as is the case for the benchmark models considered here), then we consider an off-shell decay $\chi_2 \to \chi_1 X$, where $X$ denotes some number of (unspecified) light states, with momenta $k_1$, $k_2$, etc.
We remain agnostic as to the exact model for these light states and proceed with a model-independent analysis.

Source terms \eqref{eq:source_decay} and \eqref{eq:source_upscat} are either gain or loss terms depending on whether their signs are positive or negative, respectively.
For decay $\chi_2 \to \chi_1 X$, the $\chi_1$ source terms are positive (gain) whereas the $\chi_2$ source terms are negative (loss). 
For upscattering $\chi_1 \chi_1 \to \chi_2 \chi_2$, we would have the opposite scenario. 
In our notation, the sum over matrix elements encapsulates any factors arising from averaging over spins. 
We assume that the background of decay products $X$ is negligible compared to those of $\chi_1$ and $\chi_2$ states in the late Universe and also that the population of $\chi_2$ particles is small compared to $\chi_1$, so that $f_1 \gg f_2$.

Computing these source terms is an involved process. 
Appendix~\ref{app:collision_decay} provides our derivation of the decay terms and Appendix~\ref{app:collision_upscattering} for the upscattering terms. 
Here we present the results for the two-fluid Navier-Stokes equations.
The ground-state fluid equations are
\bea \label{eq:euler1_1}
D_t\rho_1 &=& - \rho_1 \boldsymbol{\nabla} \cdot \mathbf{v}_1 
+ \Gamma_2 \rho_2 
- \frac{\rho_1^2}{2\sqrt\pi \nu_1^3} \int_0^\infty d\vrel \vrel^3 \frac{\sigma_{\rm in}(\vrel)}{m_1} \exp\left( - \frac{\vrel^2}{4\nu_1^2} \right), \\
\label{eq:euler2_1}
D_t \mathbf v_1 &=& -  \frac{\boldsymbol{\nabla} P_1}{\rho_1}  +  \mathbf{g} - \Gamma_2 \frac{\rho_2}{\rho_1} \mathbf{v}_{12} , \\
D_t e_1 &=& - \frac{P_1}{\rho_1} \boldsymbol{\nabla} \cdot \mathbf{v}_1 
+ \frac{1}{\rho_1} \boldsymbol\nabla \cdot (\tilde{\kappa}_1 \boldsymbol{\nabla}e_1) 
+ \Gamma_2 \frac{\rho_2}{\rho_1} \Big( \frac{\Delta^2}{2 m_1^2}  + \tfrac{1}{2} |\mathbf v_{12}|^2 + (e_2 - e_1) \Big)
\notag \\
&\;& \qquad 
- \frac{\rho_1}{16\sqrt\pi \nu_1^3} \int_0^\infty d\vrel \vrel^3 \left( \vrel^2 - 6\nu_1^2\right) \frac{\sigma_{\rm in}(\vrel)}{m_1} \exp\left( - \frac{\vrel^2}{4\nu_1^2} \right),
\label{eq:euler3_1}
\eea
and the excited-state fluid equations are
\bea \label{eq:euler1_2}
D_t\rho_2 &=& - \rho_2 \boldsymbol{\nabla} \cdot \mathbf{v}_2 
- \Gamma_2 \rho_2 
+ \frac{\rho_1^2}{2\sqrt\pi \nu_1^3}  \int_0^\infty d\vrel \vrel^3 \frac{\sigma_{\rm in}(\vrel)}{m_1} \exp\left( - \frac{\vrel^2}{4\nu_1^2} \right) , \\
\label{eq:euler2_2}
D_t \mathbf v_2 &=& -  \frac{\boldsymbol{\nabla} P_2}{\rho_2}  +  \mathbf{g} + \frac{\rho_1^2 \mathbf{v}_{12} }{2 \rho_2 \sqrt\pi \nu_1^3}   \int_0^\infty d\vrel \vrel^3 \frac{\sigma_{\rm in}(\vrel)}{m_1} \exp\left( - \frac{\vrel^2}{4\nu_1^2} \right) \,    ,  \\
D_t e_2 &=& - \frac{P_2}{\rho_2} \boldsymbol{\nabla} \cdot \mathbf{v}_2 + \frac{1}{\rho_2} \boldsymbol\nabla \cdot (\tilde{\kappa}_2 \boldsymbol{\nabla}e_2) 
\label{eq:euler3_2} \\
&\;& + \frac{\rho_1^2 }{ 16\sqrt\pi \rho_2 \nu_1^3} \int_0^\infty d\vrel \vrel^3 \left( \vrel^2 + 6\nu_1^2 + 4|\mathbf v_{12}|^2 - 8 \frac{\delta}{m_1} - 8 e_2 \right) \frac{\sigma_{\rm in}(\vrel)}{m_1} \exp\left( - \frac{\vrel^2}{4\nu_1^2} \right),
\notag
\eea
where $\mathbf{v}_{12} = \mathbf{v}_1 - \mathbf{v}_2$ is the relative bulk velocity, $\Gamma_2$ is the $\chi_2$ decay rate, $\Delta^2 \le \delta^2$ is the mean squared momentum of the daughter $\chi_1$ in the $\chi_2$ rest frame ($\Delta^2 \approx \delta^2 - m_\phi^2$ for on-shell decays and $\Delta^2 \approx \delta^2 - \langle s_X \rangle$ for off-shell decays where $\langle s_X \rangle$ is the averaged invariant mass squared of the decay products; see footnote~\footnotemark[1] and \cref{app:collision_decay}) and the inelastic total cross section for upscattering is $\sigma_{\rm in} = \int d\Omega \, d\sigma(\chi_1 \chi_1 \to \chi_2 \chi_2)/d\Omega$. 
The latter enters through integrals over $\vrel$, the relative velocity between ground state dark matter particles involved in scattering. 
At finite mass splitting $\delta$, there is a kinematic threshold $\vrel \ge 2 \nuloss$ for upscattering to be allowed, where
\begin{align}
    \nuloss = \sqrt{\frac{2 \delta}{m_1}}\, ,
\end{align}
dubbed the loss velocity since $\tfrac{1}{2} m_1 \nuloss^2$ is the typical kinetic energy lost per collision per particle~\cite{Essig:2018pzq}.
Therefore, integrals over the inelastic scattering cross section in the above equations can be rewritten with $2 \nuloss$ as their lower bounds.

\subsection{Effective one-fluid equations}

In the limit that $\chi_2$ is short-lived, we can reduce the two-fluid Eqs.~(\ref{eq:euler1_1}-\ref{eq:euler3_2}) to an effective one-fluid description.
To begin, we define the total mass density and mass-weighted bulk velocity and specific energy as follows
\begin{align}
    \rho &= \rho_1 + \rho_2, \\
    \mathbf{v} &= \frac{\rho_1\mathbf{v}_1 + \rho_2\mathbf{v}_2}{\rho} , \\
    e &= \frac{\rho_1e_1 + \rho_2e_2}{\rho}.
\end{align}
We apply these definitions to Eqs. (\ref{eq:euler1_1}--\ref{eq:euler3_2}), to determine the effective one-fluid description of the generally two-fluid model.

To combine the two fluids into a single effective description, we begin by assuming that the bulk velocity is dominated by the ground state fluid, such that $\mathbf{v} \approx\mathbf{v}_1$.  This is justified due to the fact that we are concerned with the regime in which $\chi_2$ is very short-lived, and thus makes up a negligible fraction of the mass density and will not affect the overall bulk velocity. Furthermore, any $\mathbf{v}_{12}$ terms (such as those in Eq.~(\ref{eq:euler2_2})) are dropped here, as the relative velocity relaxes such that $\mathbf{v}_2 \to \mathbf{v}_1$, and therefore $\mathbf{v}_{12} \to 0$. Under these assumptions, it is simple to show that the one-fluid description of the continuity and momentum gravothermal fluid equations reduces to their standard forms:
\begin{align} \label{eq:Dt_rho}
    D_t \rho &= -\rho\boldsymbol{\nabla}\cdot\mathbf{v} , \\
    D_t \mathbf{v} &= - \frac{\boldsymbol{\nabla}P}{\rho} + \mathbf{g}, \label{eq:Dt_v}
\end{align}
where the total pressure is $P = P_1 + P_2$.
The internal energy equation, however, retains additional terms:
\begin{align}
    \label{intermediate energy equation}
    \rho D_t e = -P\boldsymbol{\nabla}\cdot\mathbf{v} + \boldsymbol{\nabla}\cdot(\tilde{\kappa}_1\boldsymbol{\nabla}e_1 + \tilde{\kappa}_2\boldsymbol{\nabla}e_2) 
    + \Gamma_2\rho_2\frac{\Delta^2}{2m_1^2}\\ \notag - \frac{\rho_1^2\nuloss^2}{4\sqrt\pi \nu_1^3} \int_{2\nuloss}^\infty d\vrel \vrel^3 \frac{\sigma_{\rm in}(\vrel)}{m_1} \exp\left( - \frac{\vrel^2}{4\nu_1^2} \right) \,. 
\end{align}
 We have explicitly written the lower bound of integration as $2 \nuloss$ in order for upscattering to be kinematically allowed.

Recall that we are considering the regime in which the decay happens on a very short timescale, meaning that the mass density $\rho_2$ is negligible compared to $\rho_1$. Thus, we can approximate $\rho\approx \rho_1$. Using the results of \cref{intermediate energy equation}, we can then write
\begin{align}
D_te = & -\frac{P}{\rho}\boldsymbol{\nabla}\cdot\mathbf{v} + \frac{1}{\rho} \boldsymbol{\nabla}\cdot(\tilde{\kappa}_1\boldsymbol{\nabla}e_1 + \tilde{\kappa}_2\boldsymbol{\nabla}e_2) + \frac{\Gamma_2\rho_2}{\rho}\frac{\Delta^2}{2m_1^2}  \notag \\
& - \frac{\nu_\mathrm{loss}^2 \rho}{4\sqrt{\pi}\nu_1^3}\int_{2\nu_{\mathrm{loss}}}^{\infty}d\vrel\vrel^3\frac{\sigma_\mathrm{in}(\vrel)}{m_1}\exp\left( - \frac{\vrel^2}{4\nu_1^2} \right). 
\label{eq:Dt_e}
\end{align}

The decay term is subdominant compared to the upscattering term and can be dropped in subsequent analyses, as the heating per decay ($\Delta^2 / 2m_1^2 $) is vastly smaller than the cooling per upscatter ($\sim \delta / m_1$), yielding a ratio $\lesssim 10^{-7}$ for any decay channel. Furthermore, for all relevant benchmarks, $\delta < m_\phi$, where the on-shell expression is negative and does not apply (with decays instead proceeding through the light states of footnote~\footnotemark[1]). 

Since $\rho_2 \ll \rho_1$, the excited state contributes negligibly to the
mass-weighted quantities, so that $e \simeq e_1$, $\nu \simeq \nu_1$, and
$P \simeq P_1$. Similarly, the conductivity of each species in a
multi-species fluid scales as $\tilde\kappa_i \propto \rho_i/\rho$, so the
conduction term is dominated by $\chi_1$, giving $\tilde\kappa \approx\tilde\kappa_1$. 

Our final form for the internal
energy equation is then
\begin{align}
    D_te = -\frac{P}{\rho}\boldsymbol{\nabla}\cdot\mathbf{v} + \frac{1}{\rho} \boldsymbol{\nabla}\cdot\left( \tilde{\kappa}\boldsymbol{\nabla}e \right) - \Gamma_{\rm cool}.
    \label{eq:De_dt_simplified}
\end{align}
Comparing this to Eq.~\eqref{eq:euler3}, we can see that this is simply the standard gravothermal fluid equation for the specific energy, with an extra term
\begin{align}
\label{eq:cooling}
    \Gamma_{\rm cool} = \frac{\rho\nu_{\text{loss}}^2}{4\sqrt{\pi}\nu^3}\int_{2\nu_{\text{loss}}}^{\infty}dv_{\text{rel}}v_{\text{rel}}^3\frac{\sigma_{\text{in}}(v_{\text{rel}})}{m_1}\exp\left( - \frac{\vrel^2}{4\nu^2} \right) 
\end{align}
that represents dissipative cooling due to upscattering.
Eq.~\eqref{eq:cooling} is the specific cooling rate, while $\rho \Gamma_{\rm cool}$ is the volumetric cooling rate.\footnote{Our cooling rate agrees exactly with the specific cooling rate from Essig et al.~\cite{Essig:2018pzq} except for an overall factor of two, i.e., $\Gamma_{\rm cool}^{\rm ours} = \tfrac{1}{2} \Gamma_{\rm cool}^{\rm Essig}$.
An explanation is that Ref.~\cite{Essig:2018pzq} calculates $\Gamma_{\rm cool}$ as the energy lost {\it per collision} divided by the mean scattering time {\it per particle}, which overcounts by a factor 2.} 
Using the definition of the inelastic scattering cross section moment
\be
\langle \sigma_{\rm in}^{(n)} \rangle
= \frac{\int_{2\nuloss}^{\infty} d\vrel \, \vrel^n \sigma_{\rm in}(\vrel) \exp\left(-\tfrac{\vrel^2}{4 \nu^2}\right)}{\int_{2\nuloss}^{\infty} d\vrel \, \vrel^n \exp\left(-\tfrac{\vrel^2}{4 \nu^2}\right)}  \, , \label{eq:moment}
\ee
\cref{eq:cooling} can be written in terms of the cross section moment as 
\begin{align}
    \Gamma_{\rm cool} = \frac{2\rho \nu \nuloss^2}{\sqrt{\pi}}  \left(1 + \frac{\nu_{\rm loss}^2}{\nu^2}\right) \exp\left(-\nu_{\rm loss}^2 / \nu^2\right) \frac{\langle \sigma_{\rm in}^{(3)} \rangle}{m_1} . \label{eq:inelastic_moment}
\end{align}
Note that if $\sigma_{\rm in}$ is constant above threshold, the moments simply reduce to $\langle \sigma_{\rm in}^{(n)} \rangle = \sigma_{\rm in}$. 

Lastly, we discuss the heat conduction term. 
For purely elastic SIDM, the heat conductivity $\kappa$ encodes how the elastic scattering cross section induces heat transfer throughout the halo~\cite{Gnedin:2000ea,Balberg:2002ue}. 
The standard ansatz is to calculate $\kappa$ separately in the short mean-free-path (SMFP) and long mean-free-path (LMFP) regimes, and then combine both results with the interpolating formula
\be
\kappa = \left( \kappa_{\rm smfp}^{-1} + \kappa_{\rm lmfp}^{-1} \right)^{-1} \, ,
\ee
such that $\kappa$ is dominated by whichever conductivity is smaller~\cite{Balberg:2002ue}. 
In the SMFP regime, the heat conductivity can be calculated from kinetic theory. 
At second order in a Chapman-Enskog expansion, following Ref.~\cite{Outmezguine:2022bhq}, we have\footnote{At first order, the result is $\kappa_{\rm smfp} = b \nu/\langle \sigma_{\rm vis}^{\rm (7)}\rangle$, which is typically very close to the second-order result~\cite{Outmezguine:2022bhq}.}
\be
\kappa_{\rm smfp} = \frac{b \nu}{\langle \sigma_{\rm vis}^{\rm (eff)}\rangle }  \,, \label{eq:kappa_smfp}
\ee
where $b = 25 \sqrt{\pi}/32$ is a numerical constant and
\be
\langle\sigma_{\rm vis}^{\rm (eff)}\rangle = \frac{28 \langle\sigma_{\rm vis}^{(7)}\rangle^2 + 80 \langle\sigma_{\rm vis}^{(7)}\rangle \langle\sigma_{\rm vis}^{(11)}\rangle  - 64\langle\sigma_{\rm vis}^{(9)}\rangle^2}{77 \langle\sigma_{\rm vis}^{(7)}\rangle - 112 \langle\sigma_{\rm vis}^{(9)}\rangle + 80 \langle\sigma_{\rm vis}^{(11)}\rangle} \, . \label{eq:vis_eff_moment}
\ee
The viscosity (or heat conduction) cross section $\sigma_{\rm vis} = \int d\Omega \, \sin^2 \theta \, d\sigma/d\Omega$ is the relevant angular moment. 
The velocity moments of the viscosity cross section are
\be
\langle \sigma_{\rm vis}^{(n)} \rangle
= \frac{\int_{0}^{\infty} d\vrel \, \vrel^n \sigma_{\rm vis}(\vrel) \exp\left(-\tfrac{\vrel^2}{4 \nu^2}\right)}{\int_{0}^{\infty} d\vrel \, \vrel^n \exp\left(-\tfrac{\vrel^2}{4 \nu^2}\right)}   \label{eq:moment_vis} \, .
\ee
In the LMFP regime, where the mean interaction length becomes larger than the gravitational scale height, the conductivity can only be estimated on dimensional grounds~\cite{Balberg:2002ue} and tuned in comparison to $N$-body simulations~\cite{Ahn:2004xt,Essig:2018pzq,Nishikawa:2019lsc,Yang:2022zkd}.
Following Refs.~\cite{Yang:2022zkd,Gad-Nasr:2023gvf}, the LMFP result is
\be
\kappa_{\rm lmfp} =  \frac{9a C \nu^3 \rho}{16\pi G} \frac{\langle \sigma_{\rm vis}^{(7)}\rangle}{m_1^2} \,, \label{eq:kappa_lmfp}
\ee
where $a = 4/\sqrt{\pi}$ and $C$ is an $\mathcal{O}(1)$ numerical constant.
Values in the literature span $C \sim 0.6 - 0.8$ and in this work we fix $C=0.75$.

For dSIDM, both elastic and inelastic scattering can transfer heat throughout the halo.
In the limit that $\nuloss \ll \nu$ (but assuming $\chi_2$ is still short-lived), heat conduction is set by the {\it total} viscosity cross section over all final states 
\be
\sigma_{\rm vis} = \int d\Omega \sin^2 \theta \left( \frac{d\sigma(\chi_1 \chi_1 \to \chi_1 \chi_1)}{d\Omega} + \frac{d\sigma(\chi_1 \chi_1 \to \chi_2 \chi_2)}{d\Omega} \right) \, . \label{eq:sigma_vis_total}
\ee
In this work, for simplicity, we take Eq.~\eqref{eq:sigma_vis_total} as the viscosity cross section for evaluating $\kappa$ according to the above formulas.
In the limit $\nuloss \gg \nu$, only the elastic channel contributes to heat conduction, as expected, since the inelastic channel is Boltzmann-suppressed.

\section{Simulation results \label{sec:sim1}} 

We solve the gravothermal fluid equations for dSIDM numerically using two independent codes, \gravothermalsidm and \spherical, in order to validate our results and demonstrate the robustness of our treatment.
\gravothermalsidm~\cite{Nishikawa:2019lsc, Outmezguine:2022bhq, Gad-Nasr:2023gvf} is a publicly available code\footnote{\url{https://github.com/kboddy/GravothermalSIDM}} that solves the gravothermal fluid equations under the assumption of spherical symmetry and hydrostatic equilibrium~\cite{Pollack:2014rja}.
Here, we have modified these equations in a similar manner to Ref.~\cite{Essig:2018pzq} to account for dissipation and the system of gravothermal equations is given by
\begin{align}
    \frac{\partial M}{\partial r} &= 4 \pi r^2 \rho,  
    &\frac{\partial (\rho \nu^2)}{\partial r} &= - \frac{GM\rho}{r^2},  \\
    \frac{L}{4 \pi r^2} &= - \kappa \frac{\partial T}{\partial r}, 
    &\frac{1}{4 \pi r^2} \frac{\partial L}{\partial r} &= - \rho \nu^2 D_t \ln \left( \frac{\nu^3}{\rho} \right) - \rho \Gamma_{\rm cool} \, ,
\end{align}
where $\Gamma_{\rm cool}$ is the specific cooling rate that we derived in Eq.~\eqref{eq:inelastic_moment}. 

On the other hand, \spherical~\cite{blaff_spherical,SPHerical_paper} solves the one-fluid gravothermal equations~\eqref{eq:Dt_rho}, \eqref{eq:Dt_v}, and \eqref{eq:De_dt_simplified} directly, assuming spherical symmetry but without imposing hydrostatic equilibrium, using a spherically symmetric version of smoothed particle hydrodynamics (SPH).
This approach treats the halo as an ensemble of dynamical concentric shells, each representing a spherically averaged fluid parcel, which is far less expensive than fully three-dimensional SPH.

Both codes simulate idealized, non-cosmological halos, with initial conditions taken as Navarro-Frenk-White (NFW) profiles~\cite{Navarro:1995iw,Navarro:1996gj} in hydrostatic equilibrium, characterized by scale density and radius $(\rho_s, r_s)$.
That is, we assume the timescale for dark matter interactions is longer than the cosmological formation of halos through hierarchical collapse and mergers.
The latter yields effectively collisionless CDM halos that are then modified through subsequent particle dynamics (for isolated halos).
This assumption breaks down for some parameters we consider where core-collapse is rapid, which would necessitate a different approach where collapse and/or fragmentation occur simultaneously with halo formation. 
Here, we simply state that our work helps identify where such issues arise.

For the initial halo parameters $(\rho_s, r_s)$, we adopt values from a population of 18 dwarf and low surface brightness (LSB) galaxies, which were previously analyzed in Refs.~\cite{Kamada:2016euw, Essig:2018pzq}. 
Table~\ref{tab:halo_params} shows best-fit halo parameters fit for purely elastic SIDM with constant $\sigma/m = 3$ cm$^2$/g.
While these halo parameters were clearly fit with this specific model, in this work these parameters are meant simply to be representative of such halos and we are not performing actual fits.

\begin{table}[b]
\centering
\setlength{\tabcolsep}{2.2pt}
\begin{tabular}{ccccc|ccccc}
 Name & $M_{200}$ [$M_\odot$] & $c_{200}$ & $r_s$ [kpc] & $\rho_s$ [$M_\odot$/kpc$^{3}$] & Name & $M_{200}$ [$M_\odot$] & $c_{200}$ & $r_s$ [kpc] & $\rho_s$ [$M_\odot$/kpc$^{3}$] \\ \hline\hline
 UGC 4483 & $1.5 \times 10^{9}$ & 16.1 & 1.5 & $1.86 \times 10^{7}$ & NGC 3109 & $5.5 \times 10^{10}$ & 11.2 & 7.2 & $7.50 \times 10^{6}$ \\ \hline
 DDO 126 & $9 \times 10^{9}$ & 10.4 & 4.2 & $6.31 \times 10^{6}$ & NGC 1560 & $6 \times 10^{10}$ & 11.1 & 7.4 & $7.34 \times 10^{6}$ \\ \hline
 DDO 133 & $1.2 \times 10^{10}$ & 16.8 & 2.9 & $2.08 \times 10^{7}$ & LSB F583-1 & $8 \times 10^{10}$ & 13.9 & 6.5 & $1.28 \times 10^{7}$ \\ \hline
 DDO 154 & $1.3 \times 10^{10}$ & 14.7 & 3.4 & $1.48 \times 10^{7}$ & UGC 5750 & $8 \times 10^{10}$ & 7.4 & 12.3 & $2.72 \times 10^{6}$ \\ \hline
 NGC 2366 & $2.3 \times 10^{10}$ & 12.2 & 4.9 & $9.33 \times 10^{6}$ & UGC 3371 & $9 \times 10^{10}$ & 6.4 & 14.7 & $1.96 \times 10^{6}$ \\ \hline
 UGCA 442 & $3 \times 10^{10}$ & 11.9 & 5.5 & $8.73 \times 10^{6}$ & UGC 11707 & $10^{11}$ & 10.5 & 9.3 & $6.47 \times 10^{6}$ \\ \hline
 UGC 1281 & $3 \times 10^{10}$ & 11.9 & 5.5 & $8.73 \times 10^{6}$ & IC 2574 & $1.5 \times 10^{11}$ & 5.4 & 20.9 & $1.30 \times 10^{6}$ \\ \hline
 DDO 52 & $3 \times 10^{10}$ & 15.3 & 4.3 & $1.65 \times 10^{7}$ & UGC 5005 & $1.8 \times 10^{11}$ & 7.7 & 15.5 & $3.03 \times 10^{6}$ \\ \hline
 DDO 87 & $3.5 \times 10^{10}$ & 8 & 8.6 & $3.33 \times 10^{6}$ & UGC 128 & $3.8 \times 10^{11}$ & 9.2 & 16.6 & $4.65 \times 10^{6}$ \\ \hline
\end{tabular}
\caption{\it Halo parameters for 18 dwarf and low-surface brightness galaxies from Ref.~\cite{Kamada:2016euw}.}
\label{tab:halo_params}
\end{table}

\subsection{Benchmark particle models}

\begin{figure}[t]
    \centering
    \includegraphics[width=0.99\linewidth]{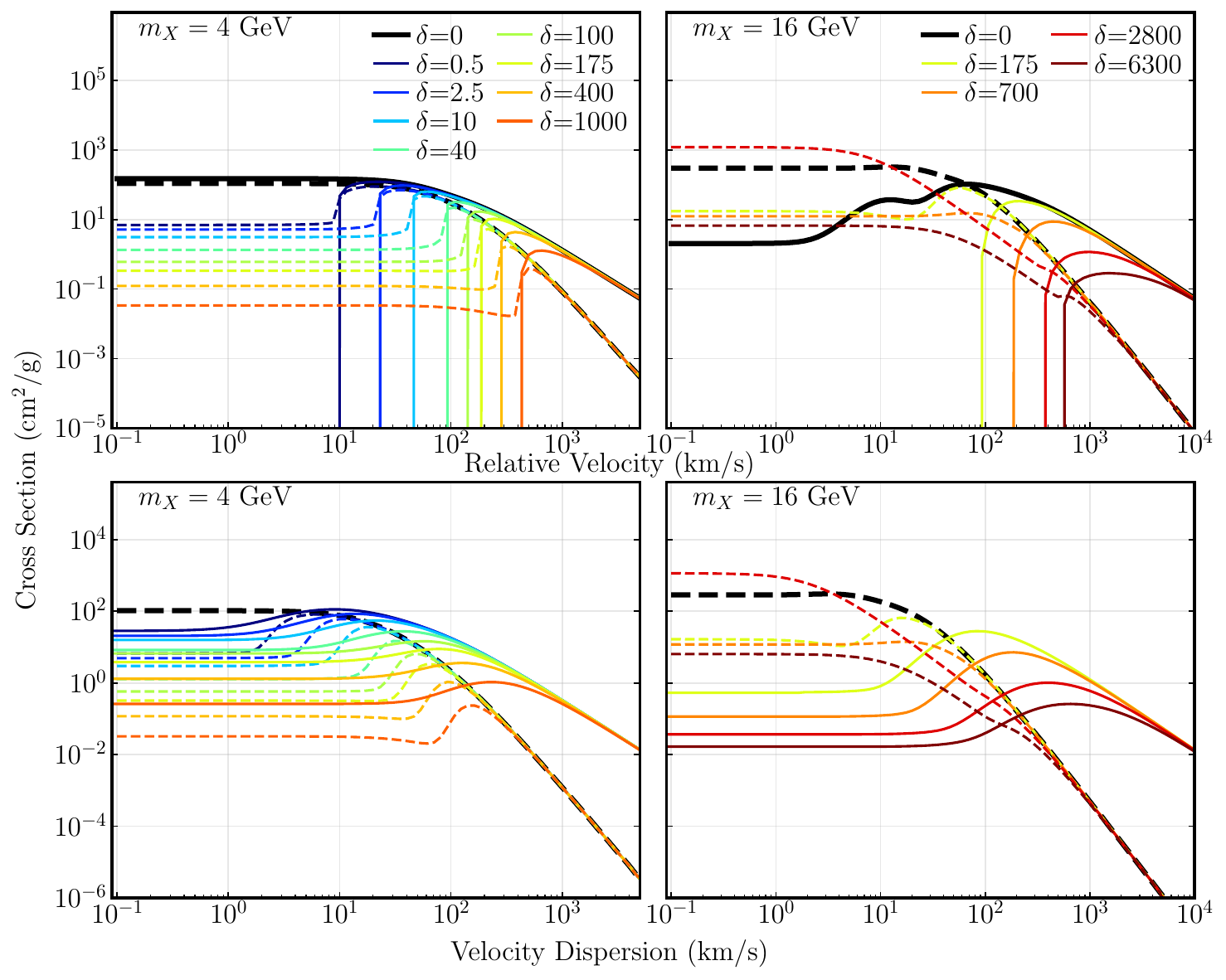}
    \\
    \caption{ \it {\bf Upper:} Inelastic (solid) and total viscosity (dashed) cross sections per unit mass as a function of the relative velocity $\vrel$, for various mass splittings $\delta$ in eV units. {\bf Lower}: The cross-section moments as a function of velocity dispersion. Solid curves represent the 3rd velocity moment of the total inelastic scattering cross-section $\langle \sigma_{\rm in}^{(3)} \rangle/m_\chi$ entering $\Gammacool$. 
    The dashed curves are the effective moment $\langle\sigma_{\rm vis}^{\rm (eff)}\rangle/m_\chi$, defined in \cref{eq:vis_eff_moment}, entering the SMFP conductivity.
    \label{fig:cross_v}}
\end{figure}

The inelastic DM model has four parameters: DM mass $m_\chi$ (previously denoted as the ground state mass $m_1$), mediator mass $m_\phi$, mass splitting $\delta$, and interaction strength $\alpha_\chi = g_\chi^2 / (4\pi)$, analogous to the fine structure constant. 
Throughout, we fix 
\begin{align}
    \alpha_\chi = 0.01 \left( \frac{m_\chi}{270 \text{ GeV}} \right) \,  ,\label{eq:alpha_relic}
\end{align}
to saturate the DM relic abundance through thermal freeze-out~\cite{Batell:2009vb,Pospelov:2007mp}.
 
For each parameter choice, we compute the velocity-dependent elastic and inelastic cross sections using two-state partial wave scattering methods, as a function of scattering velocity $\vrel$, following Refs.~\cite{Blennow:2016gde,Alvarez:2019nwt}.\footnote{In a companion paper, we will describe our quantum mechanical calculations in more detail and use them to extend our exploration of inelastic DM parameter space~\cite{quantum_team_paper}.}
From this, the required velocity moments entering \cref{eq:kappa_smfp,eq:kappa_lmfp,eq:inelastic_moment} are computed as a numerical integration over $\vrel$, as a function of $\nu$.
Finally, we input into our simulation packages interpolation functions for the tabulated cross section moments over an array of $\nu$ values.

To illustrate how inelastic scattering affects the gravothermal halo dynamics, we choose two sets of parameters, given in Table \ref{tab:benchmark_params}. 
Set A ($m_\chi = 4 \, \GeV$, $m_\phi = 1$ MeV) is representative of the perturbative Born regime, where $\alpha_\chi m_\chi / m_\phi < 1$, while Set B ($m_\chi = 16 \, \GeV$, $m_\phi = 1$ MeV) represents the nonperturbative ``resonant'' regime~\cite{Tulin:2012wi,Schutz:2014nka}.
For each set, the value of the mass splitting $\delta$ is varied to illustrate its direct effect on the elastic and upscattering cross sections.
We have chosen $m_\chi$ and $m_\phi$ such that viscosity cross sections are $\sim 0.1 - 100 \cmg$, spanning the range of values needed to solve small-scale structure issues. 
The range of $\delta$ we consider corresponds to values of $\nuloss$ spanning typical dwarf galaxy velocities, $\mathcal{O}(10) - \mathcal{O}(100) \kms$.
For all models, direct decays $\chi_2 \to \chi_1 \phi$ are kinematically forbidden, since $\delta < m_\phi$, and we assume de-excitation occurs into additional light states. 

The viscosity and inelastic scattering cross sections can be seen in the top panels of Fig.~\ref{fig:cross_v}, where the left (right) panel corresponds to $m_\chi = 4 \, \GeV$ ($16 \, \GeV$). 
The dashed and solid linestyles show the viscosity and inelastic scattering cross sections, respectively. 
The inelastic scattering cross sections have a kinematic threshold at $\vrel = 2 \nuloss$ below which inelastic scattering is forbidden. 
While the elastic channel remains open at low $\vrel$, the cross section is suppressed as $\delta$ becomes large since the process arises only through ladder diagrams with virtual excited states.
This suppression is monotonic with increasing $\delta$ in the Born regime (left), but has a nontrivial behavior in the resonant regime (right).
On the other hand, in the high-energy limit $\vrel \gg \nuloss$, the mass eigenstates become effectively degenerate and cross sections approach their $\delta = 0$ values.

The bottom panels of \cref{fig:cross_v} show moments of the cross sections as a function of $\nu$, for the same parameter points and color scheme as in the top panels.
The solid lines show the third moment of the inelastic scattering cross section $\langle \sigma_{\rm in}^{(3)} \rangle/m_\chi$ which enters the cooling rate $\Gammacool$. 
This moment is not Boltzmann-suppressed for $\nu \ll \nuloss$, due to its normalization in \cref{eq:moment}, but the suppression is explicit in \cref{eq:inelastic_moment}.
Next, the dashed lines show the effective moment $\langle \sigma^{\rm (eff)}_{\rm vis} \rangle / m_\chi$ that enters the SMFP heat conductivity (cf.~\cref{eq:kappa_smfp}), which is a combination of the 7th, 9th, and 11th moments of the viscosity cross section. 
This is the quantity responsible for the redistribution of heat throughout the halo in the later stages of collapse and is numerically similar to the 7th viscosity moment in \cref{eq:kappa_lmfp} for the LMFP heat conductivity.

\begin{table}[b]
\centering
\setlength{\tabcolsep}{10pt}
\begin{tabular}{c|ccccccccc}
\hline
\multicolumn{10}{l}{Benchmark A: $m_\chi = 4\,\mathrm{GeV}$,\; $m_\phi = 1\,\mathrm{MeV}$}\\ \hline\hline
$\delta$ [eV] & 0 & 0.5 & 2.5 & 10 & 40 & 100 & 175 & 400 & 1000 \\
$\nuloss$ [km/s] & 0 & 4.7 & 10.6 & 21.2 & 42.4 & 67.0 & 88.7 & 134.1 & 212.0 \\
\hline
\end{tabular}

\vspace{0.75em}

\begin{tabular}{c|ccccc}
\hline
\multicolumn{6}{l}{Benchmark B: $m_\chi = 16\,\mathrm{GeV}$,\; $m_\phi = 1\,\mathrm{MeV}$}\\ \hline\hline
$\delta$ [eV] & 0 & 175 & 700 & 2800 & 6300 \\
$\nuloss$ [km/s] & 0 & 44.3 & 88.7 & 177.4 & 266.0 \\
\hline
\end{tabular}
\caption{\it Benchmark inelastic dark matter parameters. The loss velocity is given by $\nuloss = \sqrt{2\delta/m_\chi}$ (in units of $c$) and all models have coupling $\alpha_\chi$ fixed by the relic density.} 
\label{tab:benchmark_params}
\end{table}

\subsection{Core evolution}

\begin{figure}[t]
\includegraphics[width=0.99\linewidth]{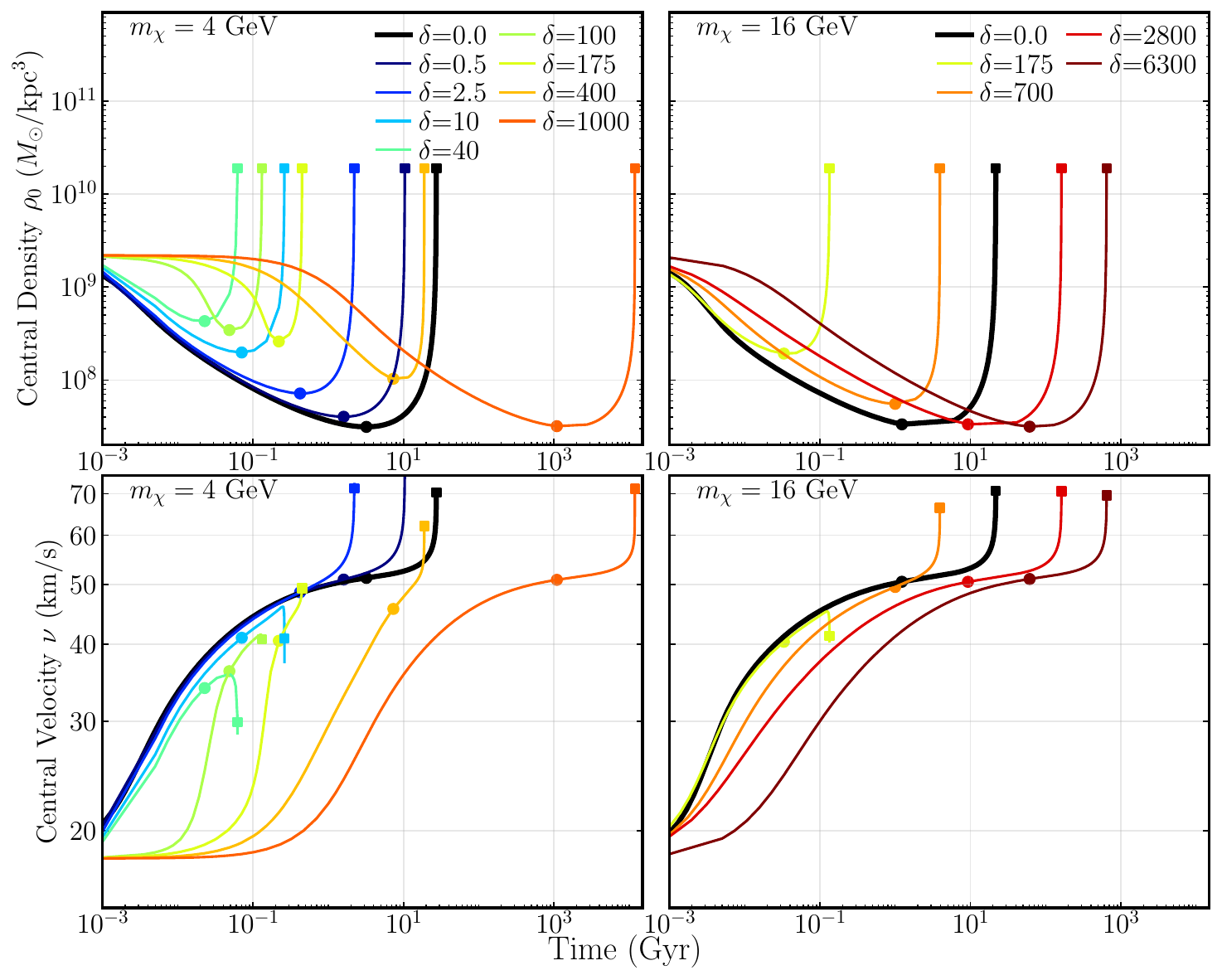}
\caption{\it  {\bf Upper:} Halo evolution of the central core density for all benchmark models in \cref{tab:benchmark_params}, where thick (black) lines are for purely elastic interactions and thin (colored) lines are for dissipative interactions.
Circles indicate points of maximal core size (minimum central density) and squares indicate core-collapse, when the central shell reaches 10$\times$ the initial density.
{\bf Lower:} Halo evolution of the central 1D velocity dispersion $\nu$ for the same models, as in the upper plots.
Results used \texttt{GravothermalSIDM} with $N=150$ shells.
}\label{fig:dens_and_vel}
\end{figure}

Using the cross section moments obtained in the previous section, we perform idealized simulations for the benchmark models in Table~\ref{tab:benchmark_params} and the initial halo parameters of LSB F583-1 (see Table~\ref{tab:halo_params}). 
Fig.~\ref{fig:dens_and_vel} shows the central density (velocity dispersion) evolution of the innermost shell in the top (bottom) panels for parameter set A (left) and set B (right). 
Circles indicate the trajectory point corresponding to the minimum core density, while squares indicate core-collapse, which we define as when the central density is $10\times$ larger than the initial density at the position of the innermost shell. 

The top panels of \cref{fig:dens_and_vel} show how the core-growing and collapse phases depend on $\delta$.
The thicker black lines show the halo evolution trajectories for elastic-only scattering with $\delta=0$, while colored lines show trajectories for inelastic halos with finite $\delta$.
As $\delta$ becomes larger, the cooling rate increases as more energy is lost per collision, leading to faster collapse, eventually reaching a minimum when $\nuloss$ is comparable to typical $\nu$ values for the halo.
For still larger values of $\delta$, the cooling rate becomes Boltzmann-suppressed as the typical particle velocities have too little kinetic energy to access the inelastic channel.
For the largest values of $\delta$ we considered, collapse times far exceed the $\delta=0$ case since elastic scattering is also suppressed, having to proceed at loop-order through off-shell excited states.

In the bottom panels of \cref{fig:dens_and_vel}, we show the central velocity dispersion over time, which shows how the halo is heated and cooled throughout its evolution.
For elastic-only SIDM, the central halo begins cold and then heats over time.
During core-collapse, although energy is flowing outward, a negative heat capacity means the halo continues to get hotter, known as the gravothermal catastrophe~\cite{LyndenBell:1968yw}.
Similar behavior is observed for finite $\delta$ as well.

However, for rapidly collapsing halos, we find that halos cool sharply as they near collapse.
Physically, if the cooling timescale is shorter than the free-fall timescale, which for a constant density core is
\be
t_{\rm ff} = \sqrt{\frac{3\pi}{32 G \rho_0}} \approx 26 \; {\rm Myr} \times \left( \frac{\rho_0}{10^{8} M_\odot / {\rm kpc}^3} \right)^{-1/2} \, ,
\ee
the inner halo will cool too rapidly to maintain hydrodynamical support and will undergo fragmentation~\cite{1977MNRAS.179..541R}.
Thus, we interpret sharp-cooling features in \cref{fig:dens_and_vel} as the onset of the fragmentation regime, for example, with regions of the halo collapsing into dark compact objects with masses and densities akin to those of stars and planets (see e.g. \cite{Buckley:2017ttd, Chang:2018bgx, Gurian:2022nbx, Bramante:2023ddr, Bramante:2024pyc,  Litterer:2025quq}).
While this effect lies beyond the scope of our spherically symmetric simulations, our results can indicate the range of particle parameters, halo evolution times, and radii where fragmentation occurs.
We leave further study of this topic to future work.

\subsection{Halo profiles}

\begin{figure}[t]
\includegraphics[width=0.99\linewidth]{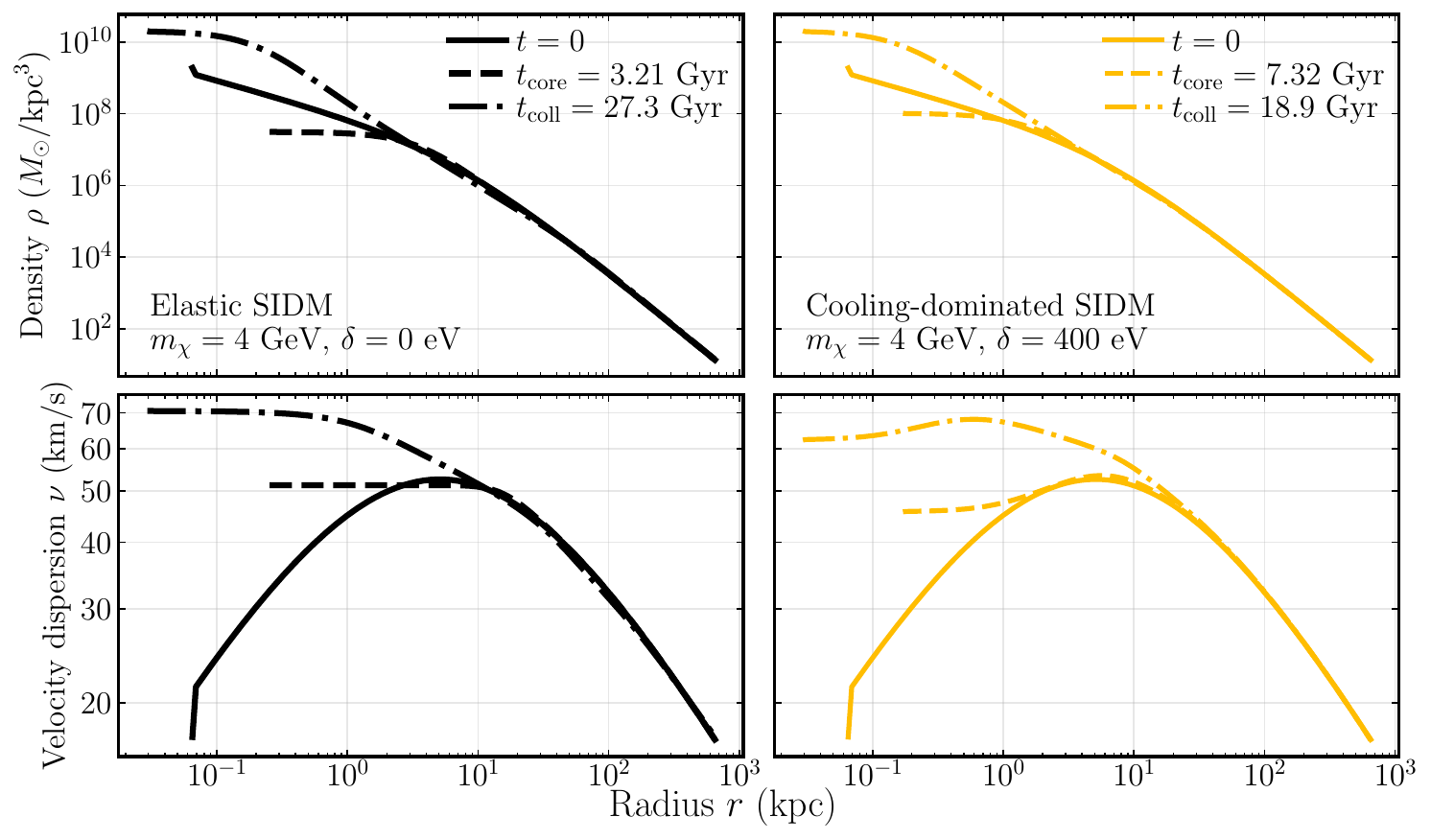}
\caption{ \it {\bf Upper:} Halo density profile as a function of $r$ for the elastic scattering-only case on the left and cooling-dominated case on the right. {\bf Lower:} Halo velocity dispersion profile as a function of $r$ for both cases.  Panels show the profiles at the initial time $t=0$, the time of maximal core size $t_{\rm core}$, and the core-collapse time $t_{\rm coll}$. The maximal core times and core-collapse times for each case are shown by circles and squares in \cref{fig:dens_and_vel}. 
Results used \texttt{GravothermalSIDM} with $N=150$ shells.}\label{fig:dens_profile}
\end{figure}

Next, we consider how radially-dependent halo profiles change over their evolution.
\cref{fig:dens_profile} shows the halo density (upper panels) and velocity dispersion (lower panels) profiles as a function of distance from the center $r$, at three different snapshots in time, for our $m_\chi = 4$ GeV model. 
The solid, dashed, and dotted-dashed curves correspond to the initial time $t=0$, the maximal core size time $t_{\rm core}$, and the core-collapse time $t_{\rm coll}$, respectively. 
The left panels are for SIDM with only elastic interactions ($\delta=0$), while the right panels are for dSIDM, with $\delta = 400$ eV.
These examples were chosen because although they collapse at a similar time, $t_{\rm coll} \approx 20~{\rm Gyr}$, they have very different elastic and inelastic cross sections:
\begin{itemize}
\item The SIDM model has $\langle\sigma_{\rm vis}^{\rm (eff)}\rangle/m_\chi \approx 10~{\rm cm^2/g}$, evaluated at the maximal core dispersion $\nu_0 = 50~ {\rm km/s}$, and $\Gammacool = 0$.
\item The dSIDM model has $\langle\sigma_{\rm vis}^{\rm (eff)}\rangle/m_\chi \approx 0.1~{\rm cm^2/g}$ and $\langle \sigma_{\rm in}^{(3)} \rangle/m_\chi \approx 2~{\rm cm^2/g}$, evaluated at the maximal core dispersion $\nu_0 = 45~ {\rm km/s}$, which yields a cooling-dominated halo.
\end{itemize}

Together, \cref{fig:dens_and_vel,fig:dens_profile} show that despite similar collapse times, the halos follow different evolutionary trajectories and the resulting profiles can differ. 
For instance, despite core-collapse taking longer for the SIDM halo, the dSIDM halo takes a longer time to reach its maximal core size due to the smaller viscosity cross section.
However, upon reaching this point, the dSIDM halo collapses faster than the SIDM halo as the former becomes strongly cooled.
For SIDM, heat is more slowly transferred out of the core, with heat conduction becoming less efficient as the halo enters the SMFP regime. 
In addition, at their maximal core size, dSIDM halos have higher central densities compared to their SIDM counterparts, even at a small mass splitting. (This is also shown neatly in \cref{fig:dens_and_vel}; circles indicate time and density of maximal core size.) 
However, despite the interplay between heat conduction and dissipation, SIDM and dSIDM density profiles during the core-collapse phase look very similar.

For the $\nu$ profiles (lower panels), the differences between the models are more visible. 
For SIDM, the inner halo progresses from a positive to an isothermal to a negative temperature gradient, corresponding to the usual core-growth, maximal-core, and core-collapse phases of gravothermal evolution.
For dSIDM, the isothermal phase is never reached before the inner density transitions into the strongly cooled regime.
At this point, cooling dominates heat conduction, which (due to a negative heat capacity) leads to an increasing velocity dispersion.

\begin{figure}[t]
\includegraphics[width=0.99\linewidth]{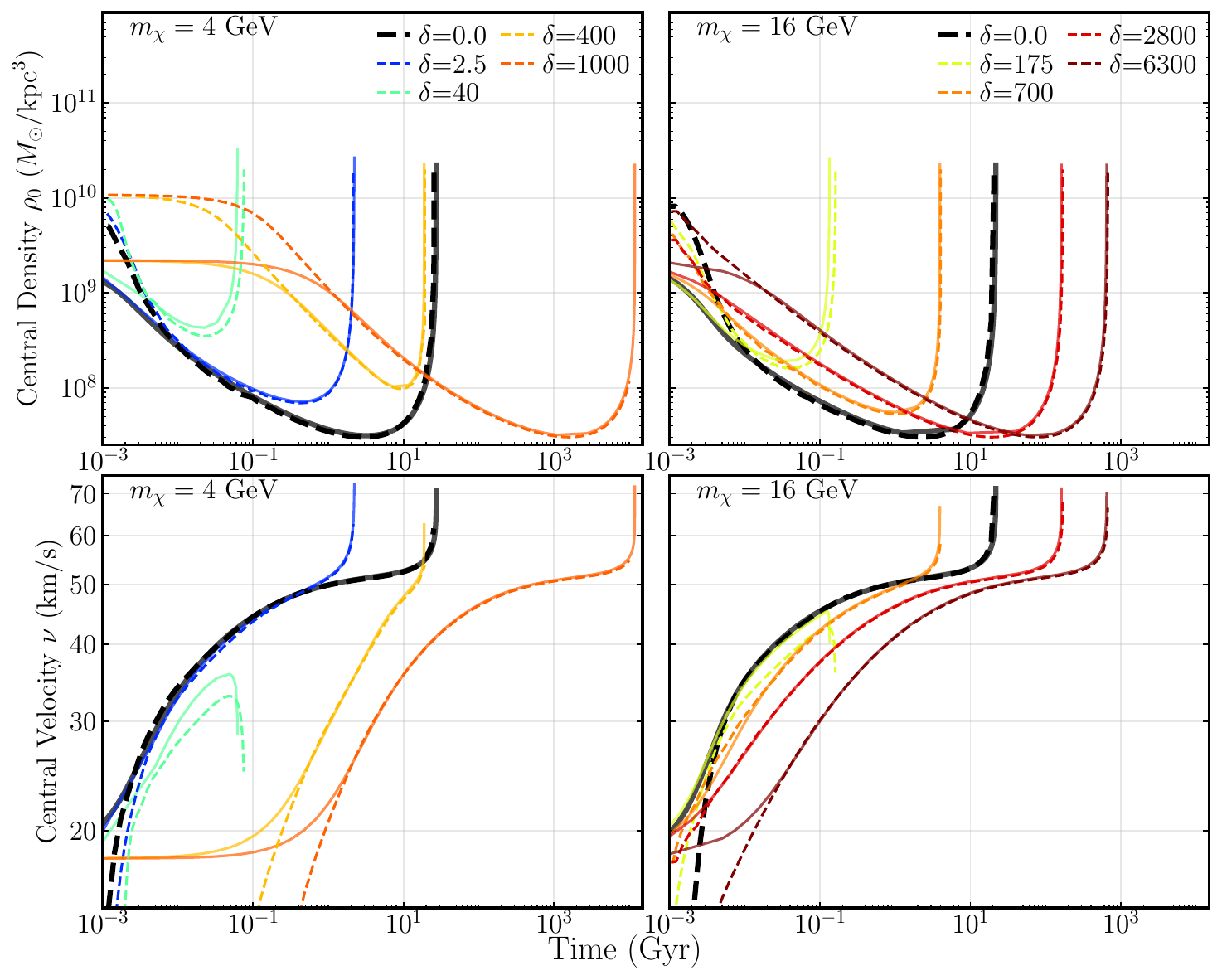}
\caption{\it Comparison between \spherical (dashed) and \gravothermalsidm (solid) codes. {\bf Upper:} central density evolution, {\bf Lower:} velocity dispersion evolution. The \spherical simulations used a minimal radius of 0.001$r_s$ and 200 shells, while the \gravothermalsidm simulations used a minimal radius of 0.01$r_s$ and 150 shells. The difference in minimal radius is the reason for the difference in starting density, but the density curves go on to align, even with the difference in shells. }
\label{fig:sim_comparison}
\end{figure}

\subsection{Comparison}

Here we compare between \gravothermalsidm and \spherical to test for robustness and convergence.
\cref{fig:sim_comparison} (similar to \cref{fig:dens_and_vel}) compares the halo evolution for \spherical runs (dashed curves) and \gravothermalsidm runs (solid curves).
Once again, benchmark values of $m_\chi$, $m_\phi$, and $\delta$ are taken from Table \ref{tab:benchmark_params}. 
Although we have chosen different minimum shell radii and numbers of shells between the simulations, we find that our results are mostly insensitive to these choices.
We see that the resulting time evolution of the central density (top panels) and velocity dispersion (bottom panels) are broadly consistent across simulations, especially for more slowly collapsing halos. 
Thus, we have demonstrated the robustness of our implementation of velocity-dependent inelastic self-interactions into the gravothermal fluid framework.

Discrepancies appear largest for systems that collapse quickest.
This disagreement may be due to different discretization errors between the codes, each with its own implementation of the gravothermal dynamics.
We have not tested convergence with $N$ for such rapidly collapsing halos since their dynamics likely lies within the fragmentation regime, which is anyway beyond the validity of a spherically symmetric analysis.
However, for these halos, we do see a sharp decrease in velocity dispersion at the onset of core-collapse in both simulation packages. 
This supports our claim that this is not a numerical or implementation artifact, but a consistent behavior from the efficient cooling that likely signals the onset of fragmentation. 


\section{Core formation and collapse timescales \label{sec:timescales}} 

For dSIDM, the evolution and collapse of DM halos depend on the underlying collisional microphysics as well as the size and structure of the halo itself.
In the gravothermal fluid framework, the latter is represented by the initial halo parameters ($\rho_s, r_s$) for the simulations, i.e., the halo profile in the absence of collisions.
The particle model here spans a four-dimensional parameter space ($\alpha_\chi, m_\chi, m_\phi, \delta$), or a three-dimensional space with the thermal relic density constraint.
With a rich variety of angular- and velocity-dependent cross sections for elastic and inelastic scattering,
exploring this entire landscape through direct simulations (even with the simplified fluid framework) is simply not feasible.

Our goal here is a semi-analytic calculation of the gravothermal collapse time $t_{\rm coll}$ for halos with both elastic and dissipative collisions.
Since many observed dwarf galaxies have cored halos, the absence of collapse can exclude much parameter space without the need for simulations or detailed analyses~\cite{Essig:2018pzq}.
These constraints were already mapped out in Ref.~\cite{Essig:2018pzq} for the case of constant cross sections.
Our work extends this argument to general models where scattering depends on both angle and velocity.

Our approach is similar in spirit to Refs.~\cite{Outmezguine:2022bhq, Gad-Nasr:2023gvf}, so first let's summarize these works.
For purely elastic SIDM, halos follow a universal time evolution trajectory, a core reaching its maximum size at time $t_{\rm core} \approx 45 \tau_{\rm el}$ and subsequently collapsing at time $t_{\rm coll} \approx 335 \tau_{\rm el}$.
Assuming most of the evolution occurs in the LMFP regime, all dependence on halo and particle parameters is absorbed into the characteristic elastic scattering timescale

\be
\label{eq:tau_el}
\tau_{\rm el} = \frac{\sqrt{\pi}}{9 C} \frac{1}{\langle\sigma_{\rm vis}^{(7)}\rangle/m_\chi} \frac{1}{\rho_{0}\nu_{0}} \, ,
\ee
where $\rho_0$ and $\nu_0$ are the central density and 1D velocity dispersion evaluated at the maximal core time.
The latter follow a universal dependence from the halo parameters, $\rho_0 \approx 2.4 \rho_s$ and $\nu_0 \approx 0.64 V_{\rm max}$, where $V_{\rm max} \approx 1.65 \sqrt{G \rho_s r_s^2}$ is the maximum circular velocity.
Angular and velocity-dependence for scattering is accounted for in the appropriate viscosity cross section moment~\eqref{eq:moment_vis}, evaluated at $\nu = \nu_0$, entering the LMFP conductivity governing most of the halo's evolution.
For a contact-type interaction with constant elastic cross section $\sigma$, we note $\langle\sigma_{\rm vis}^{(7)}\rangle = \tfrac{2}{3} \sigma$.

In this section, we first revisit the universal relations for elastic-only SIDM within our simulations.
Next, we extend the collapse time calculation to dSIDM and present a semi-analytic formula for estimating $t_{\rm coll}$ for any halo and particle parameters.
Finally, we test our collapse time estimate against direct measurement from gravothermal fluid simulations.

\subsection{Collapse time for elastic SIDM}
\label{sec:collapse_time_elastic}

\begin{figure}[t]
    \centering
    \includegraphics[width=\linewidth]{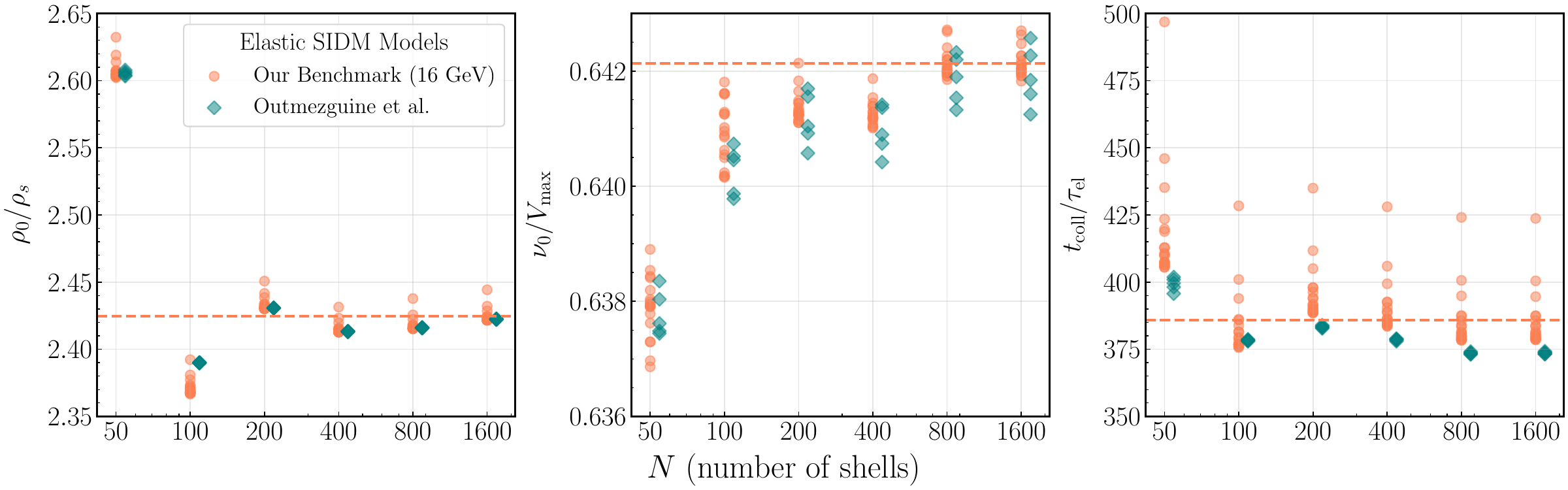}
    \caption{\it Convergence of \texttt{SPHerical} simulations with number of shells $N$ for our benchmark elastic cross section (orange circles; one model over 18 dwarf halos) and for cross section models from Outmezguine et al.~\cite{Outmezguine:2022bhq} (blue diamonds; five models over one dwarf halo).
    {\bf Left/center:} Convergence of $\rho_{0}/\rho_{s}$ and $\nu_0/V_{\rm max}$, where $\rho_0, \nu_0$ are the central density and velocity dispersion at the maximal core time.
    {\bf Right:} Convergence of the collapse time scale $t_{\rm coll}$, relative to $\tau_{el}$, the elastic scattering timescale at the maximal core time. 
    The dashed lines indicate the mean of our $N=1600$ benchmark model simulations.
    }
    \label{fig:elastic convergence}
\end{figure}

To determine the universal relations for elastic-only SIDM, we run a suite of \texttt{SPHerical} simulations across 18 dwarf and low-surface-brightness galaxies from Table~\ref{tab:halo_params}. 
The particle physics model used is our benchmark model with $m_{\chi}=16$ GeV, $m_{\phi}=1$ MeV, and $\delta=0$ for the elastic-only case. 
To test convergence, simulations vary the number of shells, from $N=50$ to $1600$ shells. 

In Figure~\ref{fig:elastic convergence}, we test the universal relations for elastic SIDM in Outmezguine et al.~\cite{Outmezguine:2022bhq}.
Each orange dot represents one simulation of our benchmark model for a different number of shells $N$, while the scatter in points at fixed $N$ corresponds to variation among the 18 initial halo parameters.
As a function of $N$, the panels show $\rho_0/\rho_s$ (left) and $\nu_0/V_{\rm max}$ (center), both at maximal core size, and $t_{\rm coll} / \tau_{\rm el}$ (right). 
The dashed lines show our converged values, taken as the mean over $N=1600$ simulations, which are
\be
\rho_0 / \rho_s \approx 2.42 \, , 
\quad
\nu_0 / V_{\rm max} \approx 0.642 \, , 
\quad
t_{\rm coll} / \tau_{\rm el} \approx 386 \, .
\label{eq:elastic_scaling_consts}
\ee
The first two values are in good agreement with Ref.~\cite{Outmezguine:2022bhq}, up to rounding, but our value of $t_{\rm coll} / \tau_{\rm el}$ is $\mathcal{O}(15\%)$ larger than their value (335).

We investigate the discrepancy in $t_{\rm coll} / \tau_{\rm el}$ through an additional suite of simulations with different particle models.
We take the same elastic cross section models from Ref.~\cite{Outmezguine:2022bhq} (runs 1-5 therein), with a single initial halo DDO 133 which has nearly identical $(\rho_s, r_s)$ to these runs.
The results are also shown in Fig.~\ref{fig:elastic convergence} (blue diamonds).
Each diamond represents one simulation with different velocity-dependent cross sections, for different number of shells $N$, though the scatter between models is indistinguishable for $N \ge 100$.
The results are in agreement with our previous simulations, which excludes differences in particle models as the explanation for the discrepancy.
We note that Ref.~\cite{Outmezguine:2022bhq} used the fifth moment $\langle \sigma_{\rm vis}^{(5)} \rangle$ in the LMFP conductivity, while we used $\langle \sigma_{\rm vis}^{(7)} \rangle$ following later work~\cite{Yang:2022zkd,Gad-Nasr:2023gvf}. 
This also cannot explain the discrepancy since their run 1 cross section is effectively velocity-independent, in which case the different moments are identical. 

\subsection{Dissipative collapse timescale}

\label{sec:dissipative_collapse_timescale}

With dissipation, the universal evolution of SIDM halos is broken by the additional timescale for cooling
\begin{align} \label{eq:tau_cool}
\tau_{\mathrm{cool}} = \frac{e_0}{\Gamma_{\rm cool}} \, ,
\end{align}
where the specific cooling rate is 
\be \label{eq:gamma_cool}
\Gamma_{\rm cool} = \frac{2 \rho_0 \nu_0 \nuloss^2 }{\sqrt \pi} \frac{\langle \sigma_{\rm in}^{(3)} \rangle}{m_\chi} \left(1 + \frac{\nuloss^2}{\nu_0^2}\right) \exp\left(-\frac{\nuloss^2}{\nu_0^2}\right) \, .
\ee
As above, $\rho_0$ and $\nu_0$ are evaluated at maximal core, $e_0 = \tfrac{3}{2} \nu_0^2$ is the specific energy in the core, and the inelastic scattering cross section moment~\eqref{eq:moment} entering the cooling rate is also evaluated at $\nu=\nu_0$.

The collapse time for a halo with both heat conduction and cooling is taken to be
\begin{align} \label{eq:t_collapse_full}
t_{\rm coll}^{-1} = t_{\rm el}^{-1} + t_{\rm cool}^{-1} \, .
\end{align}
Here, the collapse time for elastic SIDM with negligible cooling is $t_{\rm el} = A \tau_{\rm el}$, with $A \approx 386$, as in Sec.~\ref{sec:collapse_time_elastic}.
Similarly, we take the collapse time for cooling-dominated halos $t_{\rm cool}$ to be proportional to the cooling timescale~\eqref{eq:tau_cool}, writing $t_{\rm cool} = B \tau_{\rm cool}$ where $B$ is a constant (to be determined).
Eq.~\eqref{eq:t_collapse_full} assumes the two collapse times combine in their inverses, i.e., as rates, such that the collapse time is dominated by whichever time is smallest.
We also define the ratio
\be
\xi = t_{\rm coll} / t_{\rm el} \, ,
\ee
satisfying $0 < \xi \le 1$, to represent the impact of cooling~\cite{Essig:2018pzq}.
Weakly cooled halos have $\xi \to 1$ and strongly cooled halos have $\xi \to 0$.

With the additional timescale $\tau_{\rm cool}$, all halos no longer follow a common time-evolution trajectory. 
Practically speaking, there is the question of how to evaluate Eq.~\eqref{eq:gamma_cool}: what are $\rho_0$ and $\nu_0$? Can we take the same maximal core values as for purely elastic SIDM?
Fig.~\ref{fig:dens_and_vel} makes it obvious that this naive guess is doomed to fail.
Strongly cooled halos never reach the same maximal core size as weakly cooled halos.
The onset of cooling domination halts the core-growing phase prematurely and collapse ensues.
Therefore, at their maximal core size, strongly cooled halos have larger central densities $\rho_0$ and smaller dispersions $\nu_0$.
For our benchmark models, $\rho_0$ can be up to an order of magnitude larger than for elastic SIDM and $\nu_0$ is typically reduced by an $\mathcal{O}(1)$ factor.
(Although $\nu_0$ is more stable than $\rho_0$, $\tau_{\rm cool}$ is exponentially sensitive to $\nu_0$.)

\begin{figure}[t]
\includegraphics[width=1\linewidth]{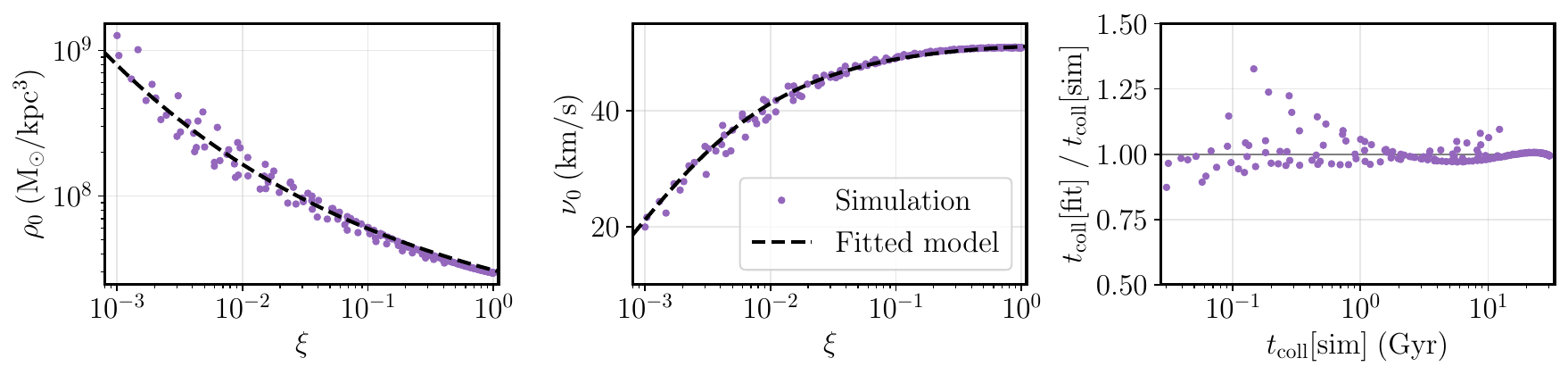}
\caption{\it Central density $\rho_0$ (left) and velocity dispersion $\nu_0$ (center) at maximal core extent are universal functions of $\xi = t_{\rm coll} / t_{\rm el}$ over 200 dSIDM simulations (one simulation=one dot).
With a parametric fit (dashed lines), we estimate the collapse time over a wide range of weakly and strongly cooled halos.
Right panel shows that $t_{\rm coll}$ from our fitted model {\rm [fit]} accurately reproduces $t_{\rm coll}$ from our simulations {\rm [sim]}, typically within $\mathcal{O}(10\%)$. \label{fig:xi_plot}}
\end{figure}

To clarify the issue, we performed an ensemble of 200 constant-cross-section simulations with
\be
\sigma_{\rm el}/m_\chi = 15 \cmg \, , 
\quad
\sigma_{\rm in}/m_\chi \in [0.5, 50] \cmg \, ,
\quad
\nuloss \in [1, 100] \kms
\ee
and LSB F583-1 halo parameters (see Table~\ref{tab:halo_params}).
With these inputs, the elastic collapse time $t_{\rm el} \approx 30 \; {\rm Gyr}$ is the same for all runs. 
The inelastic parameters span from weak to strong cooling, covering three orders of magnitude in $t_{\rm coll}$, from essentially nondissipative halos, with $t_{\rm coll} \approx t_{\rm el}$, to strongly cooled halos with $t_{\rm coll} = 30 \; {\rm Myr}$.
We also evaluate $\rho_0$ and $\nu_0$ at the maximal core time and their values span factors of $\sim 40$ and $\sim 2.5$, respectively, over the range of our simulations.
It looks like universality for SIDM halo evolution is in shambles.

Or is it? It turns out that this variation collapses to a one-dimensional function of $\xi$.
Intuitively, this makes sense: if SIDM halos follow a common time evolution, then $\xi$ serves as an effective time coordinate related to when dissipative halos depart from their universal trajectory.
Therefore, deviations from elastic-only SIDM should be a function of $\xi$ alone.

To illustrate this point, Figure~\ref{fig:xi_plot} shows the results of the simulation ensemble, each point representing one simulation from which $\rho_0$, $\nu_0$, and $t_{\rm coll}$ have been directly measured.
The left and middle panels show $\rho_0$ and $\nu_0$, respectively, as a function of $\xi$.
Most of the variation in $\rho_0$ and $\nu_0$ appears as a one-dimensional function when plotted versus $\xi$, with little scatter.
We fit the $\xi$-dependence of $\rho_0$ and $\nu_0$ according to the purely parametric ansatz
\bea \label{eq:rho_0_xi}
\rho_0(\xi) &=& \rho_{0,{\rm el}} \exp\left(a (\xi^{-p} - 1)\right) \, ,
 \\
\nu_0(\xi) &=& \nu_{0,{\rm el}} \exp\left(b (1 - \xi^{-q})\right) \, ,
\label{eq:nu_0_xi}
\eea
where $\rho_{0,{\rm el}} = 2.4 \rho_s$ and $\nu_{0,{\rm el}} = 0.64 V_{\rm max}$ are the elastic-only values. 
The parameters $a, b, p, q$ are determined through a fit, described in Appendix~\ref{sec:collapse_model_fit}, and given in Table~\ref{tab:collapse_model_parameters}. 
Taking these values, the dashed lines show our ansatz for $\rho_0$ and $\nu_0$ versus $\xi$, in good agreement with values from simulations.

\begin{table}[b]
\centering
\setlength{\tabcolsep}{6pt}
\begin{tabular}{ccccccc}
$B$ & $a$ & $b$ & $p$ & $q$ & $k$ & $\ell$ \\
\hline
1.451 & 1.205 & 0.0147 & 0.189 & 0.594 & 0.138 & -0.0293 \\
\end{tabular}
\caption{Collapse time model parameters fit in Appendix~\ref{sec:collapse_model_fit}.}
\label{tab:collapse_model_parameters}
\end{table}

Lastly, to evaluate $\tau_{\rm cool}$, we allow a bit more parametric freedom.
We allow that the characteristic density and velocity dispersion entering Eq.~\eqref{eq:tau_cool} are not exactly the maximal core values $\rho_0, \nu_0$ and instead are replaced by effective values
\bea
\label{eq:rho_eff}
\rho_0(\xi) &\to& \rho_{\rm eff}(\xi) = \rho_0(\xi) \xi^k \, , \\
\label{eq:nu_eff}
\nu_0(\xi) &\to& \nu_{\rm eff}(\xi) = \nu_0(\xi) \xi^\ell \, , 
\eea
where $k,\ell$ are additional free parameters that we fit (see Appendix~\ref{sec:collapse_model_fit}) and given in Table~\ref{tab:collapse_model_parameters}.
Physically, the cooling rate integrates over a range of densities and velocities as collapse occurs, which we model through this additional $\xi$ dependence.

Now let's summarize the collapse time evaluation and compare to the simulation ensemble.
Eq.~\eqref{eq:t_collapse_full} becomes
\be \label{eq:t_coll_xi}
t_{\rm coll} = \frac{t_{\rm el}}{1 + \frac{t_{\rm el}}{B \tau_{\rm cool}(\xi)}} ,
\ee
where $t_{\rm el} = 386 \tau_{\rm el}$ as per \cref{eq:elastic_scaling_consts} and 
\be
\tau_{\rm cool}(\xi) = \frac{3\sqrt \pi \nu_{\rm eff}(\xi)}{4 \rho_{\rm eff}(\xi) \nuloss^2 } \left(\frac{\langle \sigma_{\rm in}^{(3)} \rangle_{\nu=\nu_{\rm eff}}}{m}\right)^{-1} \left(1 + \frac{\nuloss^2}{\nu_{\rm eff}(\xi)^2}\right)^{-1} \exp\left(\frac{\nuloss^2}{\nu_{\rm eff}(\xi)^2}\right) \, ,
\ee
where $\rho_{\rm eff}(\xi)$ and $\nu_{\rm eff}(\xi)$ are defined in Eqs.~\eqref{eq:rho_eff} and \eqref{eq:nu_eff}, respectively.
However, Eq.~\eqref{eq:t_coll_xi} must be evaluated implicitly since $\xi = t_{\rm coll}/t_{\rm el}$ is an input on the right side.
That is, Eq.~\eqref{eq:t_coll_xi} can be expressed as
\be
\xi = \left(1 + \frac{t_{\rm el}}{B \tau_{\rm cool}(\xi)}\right)^{-1} \, .
\ee
It is straightforward to solve this equation iteratively, starting from $\xi=1$, until convergence is reached, which then yields $t_{\rm coll} = \xi t_{\rm el}$.

Our evaluation of $t_{\rm coll}$ is shown in Fig.~\ref{fig:xi_plot} (right panel), each dot representing one simulation from our ensemble. 
The x-axis shows $t_{\rm coll}$ directly measured from each simulation, while the y-axis shows the ratio of our model for $t_{\rm coll}$~\eqref{eq:t_coll_xi}, with parameters in Table~\ref{tab:collapse_model_parameters}, relative to the measured value.
While collapse times vary by over three orders of magnitude across different particle models that we simulated, our ansatz for $t_{\rm coll}$ models this variation rather effectively, with typically less than $\mathcal{O}(10\%)$ scatter with respect to the true value (horizontal line at unity).

Finally, we emphasize that our setup is a purely phenomenological approach.
Though it suffices for the present work, we introduced ad-hoc functional dependencies and seven fitted parameters for the cooling collapse time.
A more physical derivation of our model (or alternatives) would be desirable, but we defer that to future work.

\subsection{Comparison to simulations}

Lastly, we confront our collapse model against numerical gravothermal simulations for a range of different particle interactions and halos.
While the collapse model was fit to simulations with constant cross sections and one halo, here we demonstrate the generalization to velocity-dependent cross sections and other dwarf halos.

\begin{figure}[t]
    \centering
    \includegraphics[width=1\linewidth]{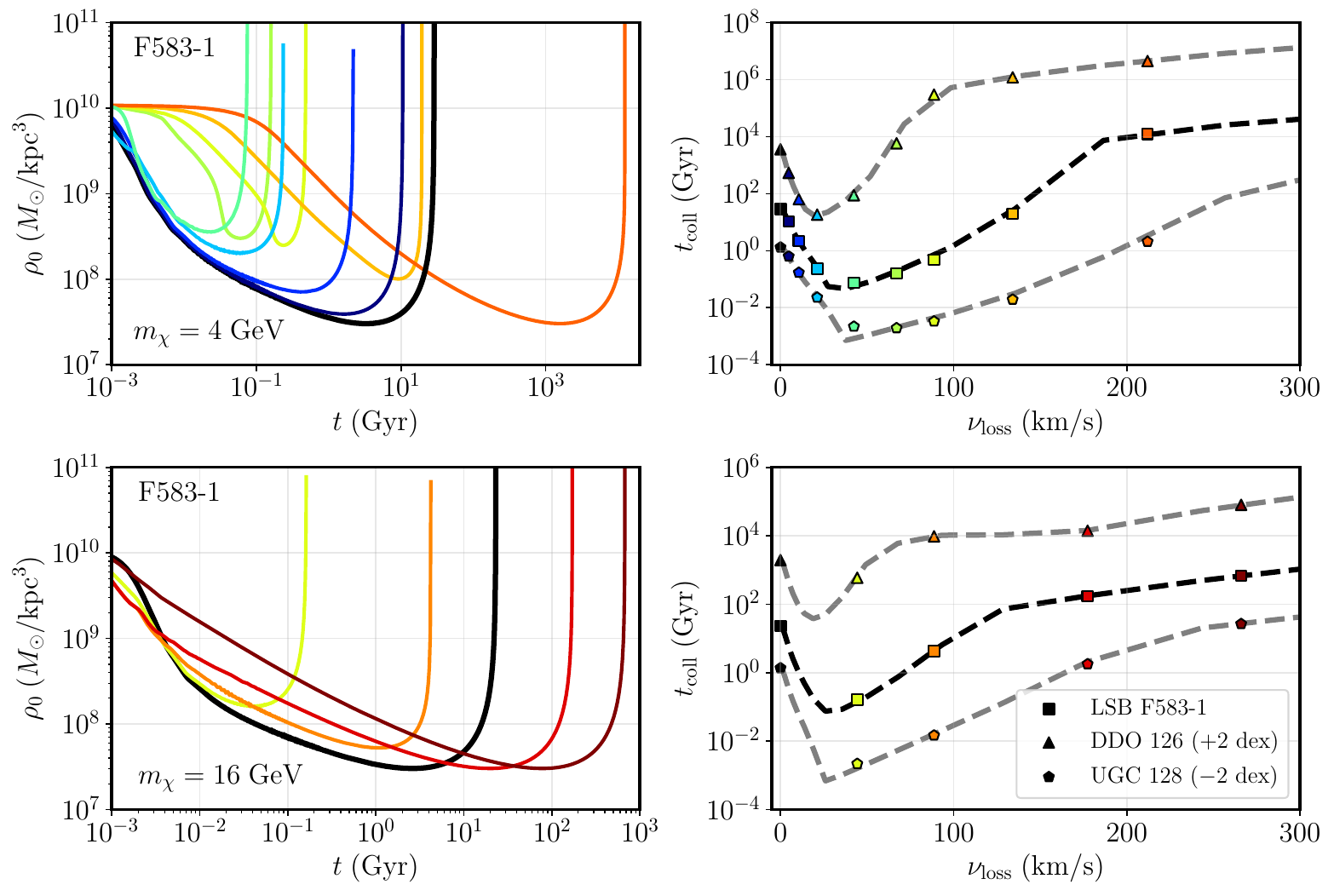}
    \caption{\it {\bf Left:} Central density evolution for benchmark models A (upper) and B (lower) for initial halo LSB F583-1, as in  \cref{fig:dens_and_vel}.
    {\bf Right}: Comparison between our semi-analytic calculations for halo collapse time $t_{\rm coll}$ (dashed curves) and values directly measured from simulations, for the same benchmark models as in the left plots and three initial halos: LSB F583-1 (squares), DDO 126 (triangles), and UGC 128 (pentagons).
    See text for further details.
    Results used the \texttt{SPHerical} code with $N=200$ shells.
    }
    \label{fig:collapse_time_scale}
\end{figure}

In Figure~\ref{fig:collapse_time_scale} (left panels), we show the central density $\rho_0$ for the LSB F583-1 halo evolving over time, for different cross sections and mass splittings for dark matter parameters listed in Table~\ref{tab:benchmark_params}, with $m_\chi = 4 \, \GeV$ in the top panel and $16 \, \GeV$ in the bottom panel. 
The thick black lines denote elastic-only scattering, with $\delta = 0$, while the thin, colored lines include cooling with $\delta \ne 0$ (color-coded as in Fig.~\ref{fig:dens_and_vel}).
The results shown here used the \spherical code with $N=200$ shells.
The right panels show the corresponding collapse times as a function of $\nuloss$.
The colored squares denote the collapse times measured directly from simulations, while the black dashed line shows the result from our collapse-time model in Sec.~\ref{sec:dissipative_collapse_timescale}.

We have performed the same set of simulations for the DDO 126 and UGC 128 halos.
Their central density evolutions (not shown) are qualitatively similar to those for LSB F583-1.
In Figure~\ref{fig:collapse_time_scale} (right panels), we show the collapse times for these halos as a function of $\nuloss$ for our models.
The colored points denote collapse times inferred from our simulations, while the gray dashed lines show the prediction of our collapse model. For clarity, results for DDO 126 (triangles) and UGC 128 (pentagons) are shifted vertically by $+2$ and $-2$ dex, respectively.
Thus, Figure~\ref{fig:collapse_time_scale} shows that our collapse time model yields a good approximation for the halo collapse time in gravothermal simulations across orders of magnitude in halo mass, mass splitting, and collapse time.

\section{Conclusions \label{sec:conclude}} 

Inelastic DM is a simple model for exploring the astrophysical phenomenology of DM self-interactions beyond the minimal paradigm of elastic-only scattering.
Here we have studied the gravothermal evolution of inelastic DM halos with velocity-dependent elastic and inelastic self-interactions mediated by a hidden vector boson.
We assumed that DM particles in halos are predominantly in the ground state, while excited states can be populated through inelastic scattering and then promptly decay back to the ground state.
These decays emit light states that escape the halo, leading to energy loss.

Starting from the two-state collisional Boltzmann equation, we derived from first principles the effective halo cooling rate from this two-step process of upscattering and decay.
Our results are applicable to both velocity-dependent and contact-type interactions.
For our numerical studies, we take velocity-dependent cross sections that have been calculated from first principles from the underlying particle parameters using a two-state quantum mechanical partial wave analysis, which will be discussed in our upcoming work~\cite{quantum_team_paper}.

Next, we performed a series of numerical simulations for idealized spherically symmetric halos within the gravothermal fluid framework for SIDM.
This approach is computationally much less expensive than $N$-body simulations, allowing for a broader exploration of different particle models and systems.
We used two different simulation packages, \gravothermalsidm and \spherical, which solve the hydrodynamical equations under different assumptions.
Both codes were modified to incorporate velocity-dependent dissipative scattering and our results were robust between them.

Consistent with previous work, we find that the gravothermal evolution of inelastic SIDM halos can be very different from that for purely elastic self-interactions, due to the finite mass splitting $\delta$.
For small $\delta$, the effect of cooling increases with $\delta$, reaching its maximal effect and giving the shortest collapse time when $\nuloss$ is comparable to the typical DM velocity dispersion.
However, for even larger $\delta$, the collapse time grows and eventually exceeds that for the purely elastic case with $\delta=0$ (with all other parameters fixed).
In this limit, inelastic transitions become Boltzmann-suppressed and even elastic scattering is suppressed since it must go through off-shell excited states, and the halo becomes effectively collisionless.
Thus, a finite mass gap $\delta$ can either accelerate or delay core-collapse compared with the $\delta=0$ case, if all other parameters are fixed.

Purely elastic SIDM halos have nearly universal evolution, set by the elastic-scattering timescale $\tau_{\rm el}$~\cite{Outmezguine:2022bhq}.
Dissipative interactions break the universal evolution by introducing an additional timescale $\tau_{\rm cool}$ for cooling.
Remarkably, however, we are able to find a semi-analytic prescription for calculating the halo collapse time $t_{\rm coll}$ with velocity-dependent elastic and dissipative interactions.
The combined effect of heat conduction and cooling on halo collapse can be described by adding the two timescales as rates, i.e., 
$t_{\rm coll}^{-1}=(A\tau_{\rm el})^{-1}+(B\tau_{\rm cool})^{-1}$, where $A, B$ are numerical constants fit from our gravothermal simulations.

However, it is also necessary to include cooling-induced changes in the core structure for calculating $\tau_{\rm cool}$. 
We find that departures from universal elastic-only evolution can be described as functions of a single parameter $\xi$ that describes whether the halo is cooled weakly ($\xi \to 1$) or strongly ($\xi \to 0$).
We introduced a parametric model to describe the $\xi$-dependent central density and velocity dispersion of halos. The model is calibrated and fit from an ensemble of gravothermal fluid simulations with contact-type interactions (see Appendix~\ref{sec:collapse_model_fit}) and reproduces the collapse times of this ensemble to $\mathcal{O}(10\%)$ across more than three orders of magnitude.
It also provides a good approximation for the velocity-dependent interactions and additional halos considered here. 
Once calibrated, this prescription provides an estimate of the collapse time directly from the initial halo properties and particle physics parameters, eliminating the need for a separate gravothermal simulation for each parameter choice.

There are several extensions of this work which would be interesting to study. 
In a companion paper \cite{constraints_team_paper}, we demonstrate how these modified collapse times can be used to constrain the microphysical parameters of the underlying inelastic DM model. 
We also aim to clarify the direct detection prospects of our model, including compatibility with the potential nuclear recoil event at LZ~\cite{LZ:2026axp}. 
Furthermore, we wish to study the combined gravothermal evolution and fragmentation phenomenology of dSIDM halos.

Lastly, by assuming the excited state has a short lifetime and negligible population in halos, we have focused on the endothermic scenario for inelastic dark matter. 
If excited states are long-lived (in comparison with other dynamical timescales), this complicates the gravothermal fluid analysis. 
The two-fluid equations would need to include downscattering (as well as other possible effects) and the effective one-fluid description would need to be revisited. 
We leave the investigation of this general scenario for a future study.

\acknowledgements

This work has emerged as a follow-up from EXPLORE (“EXPeriential Learning Opportunity through Research and Exchange”), an international innovative teaching and research program involving York University, the University of Alberta, and Goethe University Frankfurt. 
We thank all EXPLORE mentors, junior mentors, and students for stimulating discussions and fruitful collaboration.
This work was supported in part by the Natural Sciences and Engineering Research Council of Canada (NSERC). 
Our numerical calculations were enabled in part by support provided by Compute Ontario and the Digital Research Alliance of Canada (\href{http://alliancecan.ca}{alliancecan.ca}). 

{\it AI use:} We acknowledge the use of Claude Opus 5 in identifying the functional ansatz used in \cref{eq:rho_0_xi} and \cref{eq:nu_0_xi}, while all related numerical fits and analyses were performed by the authors.

\bibliography{mega_dm_bib}

\appendix

\section{Gain and loss terms due to decay}
\label{app:collision_decay}

In the fluid prescription, we define the standard quantities for the fluid density, bulk velocity, and specific internal energy for species $i$ as 
\begin{align}
    \rho_i &= \int \frac{d^3\mathbf{q}}{(2\pi)^3} m_i f_i(\mathbf{q}), \\
    \mathbf{v}_i &= \frac{1}{n_i} \int \frac{d^3\mathbf{q}}{(2\pi)^3} \frac{\mathbf{q}}{m_i} f_i(\mathbf{q}) , \\
    e_i &= \frac{1}{\rho_i} \int \frac{d^3\mathbf{q}}{(2\pi)^3} \frac{1}{m_i} \frac{|\mathbf{q} - m_i \mathbf{v}_i|^2 }{2}f_i(\mathbf{q}) \, .
\end{align}
For evaluating Boltzmann collision terms, we assume a Maxwell-Boltzmann distribution function with velocity dispersion $\nu_i$ so that
\begin{align}
    f_i(\mathbf{q}) = \frac{(2\pi)^{3/2} n_i}{m_i^3 \nu_i^3 } \exp\left( - \frac{|\mathbf{q} - m_i \mathbf{v}_i|^2}{2m_i^2 \nu_i^2} \right).
\end{align}
The simpler of the two processes is decay. 
In order to find the gain and loss terms for decay, given in \cref{eq:euler1_general_i,eq:euler2_general_i,eq:euler3_general_i}, we compute the relevant collision integrals~\cite{Kolb:1990vq} for $\chi_1$
\begin{align}
    \mathcal{S}_{\mathrm{decay},1} &= + m_1 \int d \mathcal{V}   , \label{eq:dens_decay_1_start} \\
    \bm{\mathcal{S}}_{\mathrm{decay},1} &= + \frac{1}{\rho_1}\int d \mathcal{V}~ (\mathbf{p} - m_1 \mathbf{v}_1 ) , \label{eq:velocity_decay_1_start} \\
    \tilde{\mathcal{S}}_{\mathrm{decay},1} &= + \frac{1}{\rho_1} \int d \mathcal{V} ~ m_1  \left(\frac{|\mathbf{p} - m_1 \mathbf{v}_1|^2}{2m_1^2} - e_1 \right) , \label{eq:energy_decay_1_start}
\end{align}
and for $\chi_2$
\begin{align}
    \mathcal{S}_{\mathrm{decay},2} &= - m_2 \int d \mathcal{V}, \label{eq:dens_decay_2_start} \\
    \bm{\mathcal{S}}_{\mathrm{decay},2} &= - \frac{1}{\rho_2}\int d \mathcal{V} ~ (\mathbf{q} - m_2 \mathbf{v}_2) , \label{eq:velocity_decay_2_start} \\
    \tilde{\mathcal{S}}_{\mathrm{decay},2} &= - \frac{1}{\rho_2} \int d \mathcal{V}~ m_2 \left(\frac{|\mathbf{q} - m_2 \mathbf{v}_2|^2}{2m_2^2} - e_2 \right) . \label{eq:energy_decay_2_start}
\end{align}
For simplicity, we have defined
\begin{align}
    d \mathcal{V} = d\Pi_p d\Pi_q \prod_j d\Pi_{k_j} (2 \pi)^4 \delta^4\left(q - p - \sum_j k_j \right) \left( \sum |\mathcal{M}|^2 \right) f_2 (q) \,.
\end{align}

\subsection{Decay rate}
Here we derive the decay rate for the de-excitation process $\chi_2 \to \chi_1  X$, where $X = \sum_j X_j$ is the total number of decay products with index $j$. The decay rate $\Gamma$ is given by
\begin{align}
    \Gamma = \frac{1}{2 E_q} \int \frac{d^3 \mathbf{p}}{(2\pi)^3 2E_p } \prod_j \frac{d^3 \mathbf{k}_j}{(2\pi)^3 2E_{k_j}} (2\pi)^4 \delta^4 \left(q - p - \sum_j k_j \right) \left( \sum |\mathcal{M}|^2 \right). \label{eq:decay_rate_definition}
\end{align}
In the rest frame of $\chi_2$, we have that $E_q = m_2$, and so we define the total decay rate
\begin{align}
    \Gamma_2 = \frac{1}{2m_2} \int \frac{d^3 \mathbf{p}}{(2\pi)^3 2E_p } \prod_j \frac{d^3 \mathbf{k}_j}{(2\pi)^3 2E_{k_j}} (2\pi)^4 \delta^4 \left(q - p - \sum_j k_j \right) \left( \sum |\mathcal{M}|^2 \right). \label{eq:gamma_id}
\end{align}
Note that we have assumed that the $\sum |\mathcal{M}|^2$ factor contains any symmetry factors arising from indistinguishable particles. Relating the two rates,
\begin{align}
    \Gamma = \frac{m_2}{E_q} \Gamma_2 = \frac{1}{\gamma_q} \Gamma_2.
\end{align}
In the non-relativistic approximation, $\gamma \approx 1$ and so this reduces to 
\begin{align}
    \Gamma \approx \Gamma_2.
\end{align}
In the following sections for the decay gain and loss terms, we make the substitution defined in Eq. \eqref{eq:gamma_id} for the rest-frame decay rate $\Gamma_2$. 

\subsection{Decay gain term for the ground state}

\subsubsection{Density}

The first quantity that we wish to compute is the gain term for the density of $\chi_1$ particles due to decay of $\chi_2$ particles in Eq. \eqref{eq:dens_decay_1_start}, which enters the continuity equation as
\begin{align}
    \mathcal{S}_{\mathrm{decay},1} &= \int \frac{d^3 \mathbf{p}}{(2\pi)^3 2 E_p} \frac{d^3 \mathbf{q}}{(2\pi)^3 2 E_q} \prod_{j} \frac{d^3 \mathbf{k}_j}{(2\pi)^3 2E_{k_j}} (2 \pi)^4 \delta^4\left(q - p - \sum_j k_j\right) \left( \sum |\mathcal{M}|^2 \right) m_1 f_2 (q) .
\end{align}
Rearranging the order of integration, we see that the integral over $d^3 \mathbf{p}$ and the $d^3 \mathbf{k}_j$ is the expression for the decay rate Eq.~\eqref{eq:gamma_id}, so that
\begin{align}
    \mathcal{S}_{\mathrm{decay},1} =  \int \frac{d^3 \mathbf{q}}{(2\pi)^3} \frac{m_2}{E_q} \Gamma_2 m_1 f_2(q) = \int \frac{d^3 \mathbf{q}}{(2\pi)^3  } \frac{\Gamma_2}{\gamma_q} m_1 f_2(q).
\end{align}
Then, taking $f_2(q)$ to be the Maxwell-Boltzmann distribution for $\chi_2$ particles along with a non-relativistic assumption so that $\gamma_q \approx 1$ gives 
\begin{align}
    \mathcal{S}_{\mathrm{decay},1} = m_1 n_2 \Gamma_2.
\end{align}
This lets us write the gain term due to decay for the continuity equation for $\chi_1$ as
\begin{align}
    \mathcal{S}_{\mathrm{decay},1} =  \frac{m_1}{m_2} \rho_2 \Gamma_2,
\end{align}
where taking $m_1/m_2 \approx 1$ gives us 
\begin{align}
    \boxed{\mathcal{S}_{\mathrm{decay},1} =  \rho_2 \Gamma_2}. \label{eq:source_decay_1_deriv}
\end{align}

\subsubsection{Velocity}

Next, for the bulk velocity $\mathbf{v}_1$, we have that the higher moment of the Boltzmann equation is given by Eq.~\eqref{eq:velocity_decay_1_start}. After shifting all velocities by $\mathbf{v}_2$, we have that in the non-relativistic limit where $\mathbf{p} = m_1 \mathbf{u}$,
\begin{align}
    \bm{\mathcal{S}}_{\mathrm{decay},1 } &= \frac{1}{\rho_1}  \int \frac{d^3 \mathbf{p}}{(2\pi)^3 2 E_p} \frac{d^3 \mathbf{q}}{(2\pi)^3 2 E_q} \prod_j \frac{d^3 \mathbf{k}_j}{(2\pi)^3 2E_{k_j}} (2 \pi)^4 \delta^4 \left(q - p - \sum_j k_j \right) \left( \sum |\mathcal{M}|^2 \right) \nonumber \\
    &\qquad \qquad \times (\mathbf{p} - m_1 \mathbf{v}_{12}) f_2 (q + m_2 \mathbf{v}_2),\label{eq:momentum_source_1}
\end{align}
where $\mathbf{v}_{12} = \mathbf{v}_1 - \mathbf{v}_2$. The second term is a constant with respect to $\mathbf{q}$, and so computing it is similar in procedure as $\mathcal{S}_{\mathrm{decay},1}$. To compute the first term, we define the average momentum of particle $\chi_1$ given particle $\chi_2$ has momentum $\mathbf{q}$ as
\begin{align}
    \langle \mathbf{p} \rangle_\mathbf{q} &= \frac{\int \frac{d^3 \mathbf{p}}{(2\pi)^3 2 E_p} \prod_j \frac{d^3 \mathbf{k}_j}{(2\pi)^3 2E_{k_j}} (2 \pi)^4 \delta^4(q - p - \sum_j k_j) \left( \sum |\mathcal{M}|^2 \right) \mathbf{p} f_2 (q)}{ \int \frac{d^3 \mathbf{p}}{(2\pi)^3 2 E_p} \prod_j \frac{d^3 \mathbf{k}_j}{(2\pi)^3 2E_{k_j}} (2 \pi)^4 \delta^4(q - p - \sum_j k_j) \left( \sum |\mathcal{M}|^2 \right) f_2 (q)} \nonumber \\
    &= \frac{1}{2 m_2 \Gamma_2} \int \frac{d^3 \mathbf{p}}{(2\pi)^3 2 E_p} \prod_j \frac{d^3 \mathbf{k}_j}{(2\pi)^3 2E_{k_j}} (2 \pi)^4 \delta^4 \left(q - p - \sum_j k_j \right) \left( \sum |\mathcal{M}|^2 \right) \mathbf{p}.
\end{align}
Rewriting Eq. \eqref{eq:momentum_source_1}, we have that
\begin{align}
    \bm{\mathcal{S}}_{\mathrm{decay},1} = \frac{1}{\rho_1} \int \frac{d^3\mathbf{q}}{(2\pi)^3 } (\langle \mathbf{p} \rangle_\mathbf{q} - m_1 \mathbf{v}_{12} ) \Gamma_2 f_2(q + m_2 \mathbf{v}_2).
\end{align}
To compute $\langle \mathbf{p} \rangle_\mathbf{q}$, we use the fact that momentum transformation between the rest frame of $\chi_2$ and the lab frame can be written as
\begin{align}
    \mathbf{p} = \mathbf{p}^*_\perp + \gamma_{\mathbf{q}} \left(\mathbf{p}^*_{\parallel} + \frac{\mathbf{q}}{E_q} E_p^* \right), \label{eq:momentum_Lorentz_transform}
\end{align}
where $\mathbf{p}^*_\perp$ and $\mathbf{p}^*_\parallel$ are the perpendicular and parallel components of the momentum, and $E_p^*$ is the energy of $\chi_1$ in the rest frame of $\chi_2$, where $E_p^*$ is given by
\begin{align}
    E_p^* = \frac{m_1^2 + m_2^2 - s_X}{2m_2} , \label{eq:E_p_star}
\end{align}
where we have defined
\begin{align}
    s_X = \left( \sum_j k_j \right)^2, 
\end{align}
for the invariant mass of the final (on-shell) decay products. We note that $\langle \mathbf{p}^*_\parallel \rangle = \langle \mathbf{p}^*_\perp \rangle =  0$ for isotropic decays, averaging over all angles. Thus 
\begin{align}
    \langle \mathbf{p} \rangle_\mathbf{q} = \frac{\langle E_p^* \rangle_\mathbf{q}}{m_2} \mathbf{q} ,
\end{align}
where we have taken $\gamma_\mathbf{q} = E_q/m_2$. Therefore we have
\begin{align}
    \bm{\mathcal{S}}_{\mathrm{decay},1} =  \frac{\Gamma_2}{\rho_1} \int \frac{d^3\mathbf{q}}{(2\pi)^3} \left(\frac{\langle E_p^* \rangle_\mathbf{q}}{m_2}\mathbf{q} - m_1 \mathbf{v}_{12} \right) f_2(q + m_2 \mathbf{v}_2),
\end{align}
where we now use the fact that integrating odd functions in $\mathbf{q}$ with the displaced Maxwell-Boltzmann distribution $f_2 (q + m_2 \mathbf{v}_2)$ vanishes. Therefore the first term in the brackets evaluates to zero, and $\bm{\mathcal{S}}_{\mathrm{decay},1}$ is given by
\begin{align}
    \bm{\mathcal{S}}_{\mathrm{decay}, 1} = -\Gamma_2 \frac{m_1}{m_2} \frac{\rho_2}{\rho_1} \mathbf{v}_{12}.
\end{align}
Taking $m_1/m_2 \approx 1$, we get
\begin{align}
    \boxed{\bm{\mathcal{S}}_{\mathrm{decay}, 1} = -\Gamma_2 \frac{\rho_2}{\rho_1} \mathbf{v}_{12}}.
\end{align}

\subsubsection{Energy}

The starting point for the evolution of the internal specific energy $e_1$ is from Eq. \eqref{eq:energy_decay_1_start}
\begin{align}
    \tilde{\mathcal{S}}_{\mathrm{decay},1} &= \frac{1}{ \rho_1} \int \frac{d^3 \mathbf{p}}{(2\pi)^3 2 E_p} \frac{d^3 \mathbf{q}}{(2\pi)^3 2 E_q} \prod \frac{d^3 \mathbf{k}_j}{(2\pi)^3 2E_{k_j}} (2 \pi)^4 \delta^4\left(q - p - \sum_j k_j \right) \left( \sum |\mathcal{M}|^2 \right) \nonumber \\
    &\qquad \qquad \times \left(  \frac{|\mathbf{p}- m_1 \mathbf{v}_1|^2}{2m_1^2}  - e_1\right) m_1 f_2 (q) .
\end{align}
Then, as before, we shift all velocities by $\mathbf{v}_2$, so that
\begin{align}
    \tilde{\mathcal{S}}_{\mathrm{decay},1} &= \frac{1}{m_1 \rho_1} \int \frac{d^3 \mathbf{p}}{(2\pi)^3 2 E_p} \frac{d^3 \mathbf{q}}{(2\pi)^3 2 E_q} \prod_j \frac{d^3 \mathbf{k}_j}{(2\pi)^3 2E_{k_j}} (2 \pi)^4 \delta^4 \left(q - p - \sum_j k_j  \right) \left( \sum |\mathcal{M}|^2 \right) \nonumber \\
    &\qquad \qquad \times \left(  \frac{|\mathbf{p} - m_1 \mathbf{v}_{12}|^2}{2}  - m_1^2 e_1\right) f_2 (q + m_2 \mathbf{v}_2) . \label{eq:S_decay_mid}
\end{align}
Then note that
\begin{align}
    \frac{|\mathbf{p} - m_1 \mathbf{v}_{12}|^2}{2} = \frac{| \mathbf{p}|^2}{2} + \frac{m_1^2 | \mathbf{v}_{12}|^2}{2} - m_1 \mathbf{p} \cdot \mathbf{v}_{12}.
\end{align}
We see that the last term is linear in $\mathbf{p}$, and therefore vanishes due to similar arguments as before. In a similar manner, we write $\langle |\mathbf{p}|^2 \rangle_\mathbf{q}$ by using the Lorentz transformation for momentum in Eq. \eqref{eq:momentum_Lorentz_transform} so that
\begin{align}
    \langle | \mathbf{p} |^2 \rangle_\mathbf{q} \approx \langle | \mathbf{p}^*|^2 \rangle_\mathbf{q}  + \frac{|\mathbf{q}|^2}{m_2^2} \left( \langle (E_p^*)^2 \rangle_\mathbf{q} + \langle | \mathbf{p}_\parallel^*|^2 \rangle_\mathbf{q} \right) + 2 \frac{1}{m_2} \langle E_p^* \mathbf{p}^* \rangle_\mathbf{q} \cdot \mathbf{q},
\end{align}
in the non-relativistic limit. The last term is zero due to the assumed isotropy. Substituting this expression back into \cref{eq:S_decay_mid} gives
\begin{align}
    \tilde{\mathcal{S}}_{\mathrm{decay},1} &= \frac{\Gamma_2}{m_1 \rho_1} \int \frac{d^3 \mathbf{q}}{(2\pi)^3}  \left(  \frac{\langle | \mathbf{p}^*|^2\rangle_\mathbf{q}}{2}  + \frac{|\mathbf{q}|^2}{2m_2^2} \left( \langle(E_p^*)^2 \rangle_\mathbf{q} + \langle | \mathbf{p}_\parallel^*|^2 \rangle_\mathbf{q}  \right) + \frac{m_1^2 | \mathbf{v}_{12} |^2}{2}  - m_1^2 e_1\right) f_2 (q + m_2 \mathbf{v}_2) .
\end{align}
Let us now go term-by-term to evaluate this expression. We define
\begin{align}
    \Delta^2 \equiv \langle |\mathbf{p}^*|^2 \rangle_{\mathbf{q}} = \frac{1}{2m_2 \Gamma_2} \int \frac{d^3 \mathbf{p}}{(2\pi)^3 2E_p } \prod_j \frac{d^3 \mathbf{k}_j}{(2\pi)^3 2E_{k_j}} (2\pi)^4 \delta^4 \left(q - p - \sum_j k_j \right) \left( \sum |\mathcal{M}|^2 \right) |\mathbf{p}^*|^2.
\end{align}
Using the fact that $|\mathbf{p}^*|^2 = (E_p^*)^2 - m_1^2 = (E_p^*+m_1)(E_p^* - m_1)$ along with \cref{eq:E_p_star}, we have
\begin{align}
    |\mathbf{p}^*|^2 &= (\delta^2 - s_X) \frac{(m_1+m_2)^2- s_X}{4m_2^2},
\end{align}
where we have substituted $\delta = m_2 - m_1$. In the small $\delta$-limit, we have again that $m_1 \approx m_2$, and since $s_X \leq \delta^2$ for decay to be kinematically allowed, we can approximately write $|\mathbf{p}^*|^2 \approx \delta^2 - s_X$. 
Therefore the averaging over the particle momentum gives us
\begin{align}
    \Delta^2 \approx \delta^2 - \langle s_X \rangle_\mathbf{q}.
\end{align}
For the second term with $\langle (E_p^*)^2 \rangle_\mathbf{q}$, rearranging $|\mathbf{p}^*|^2 = (E_p^*)^2 - m_1^2$ and using the definition of $\Delta$ after averaging, we get
\begin{align}
    \langle (E_p^*)^2 \rangle_\mathbf{q} = m_1^2 + \Delta^2 .
\end{align}
The third term with $\langle | \mathbf{p}_\parallel^*|^2 \rangle$ can be solved for using isotropy arguments again, where 
\begin{align}
    \langle | \mathbf{p}_\parallel^*|^2 \rangle = \langle |\mathbf{p}^*|^2 \cos^2 \theta \rangle = \frac{\Delta^2}{3},
\end{align}
where $\langle \cos^2 \theta \rangle = 1/3$ for isotropic decays. These are constants which can be taken out of the integration because they only depend on the distribution of $s_X$, and are thus independent of $\mathbf{q}$. Moreover, the $\Delta$ term can be neglected since this means that the overall term integrated with $|\mathbf{q}|^2$ will be proportional to $m_1^2 + 4\Delta^2/3 \approx m_1^2$ since $\Delta \ll m_1$. The last two terms only have constant prefactors. After substitution and using the fluid quantity definitions defined at the beginning of this Appendix, we get 
\begin{align}
    \boxed{\tilde{\mathcal{S}}_{\mathrm{decay},1} = \Gamma_2 \frac{\rho_2}{\rho_1} \left(  \frac{\Delta^2}{2m_1^2}  + \frac{| \mathbf{v}_{12}|^2}{2}  + (e_2 - e_1)\right)}. 
\end{align}
Note here that in the case that the decay is directly to $\phi$ on-shell without the existence of any lighter states, then $\langle s_X \rangle_\mathbf{q} = m_\phi^2$, and so $\Delta ^2 = \delta^2 - m_\phi^2$. Since under the mass hierarchy we assume, the decay must be through an off-shell $\phi$, one must determine $\langle s_X \rangle_\mathbf{q}$. However, as shown in the main text, this term is not important anyways since it falls out of the fluid equations at leading order.

\subsection{Decay loss term for the excited state}

Starting with Eq. \eqref{eq:dens_decay_2_start} gives us a similar process to $\chi_1$, where we get that
\begin{align}
    \boxed{\mathcal{S}_{\mathrm{decay},2} = -\rho_2 \Gamma_2},
\end{align}
which is the same as Eq. \eqref{eq:source_decay_1_deriv} but with a negative sign as this is a loss term for $\rho_2$. For the bulk velocity and specific internal energy evolution, we note that if we shift all velocities by $\mathbf{v}_2$, then these terms evaluate to zero. For the bulk velocity source term, we get
\begin{align}
    \bm{\mathcal{S}}_{\mathrm{decay},2} &= - \frac{1}{\rho_2} \int d\Pi_p d\Pi_q \prod_j d\Pi_{k_j} (2 \pi)^4 \delta^4\left(q - p - \sum_j k_j \right) \left( \sum |\mathcal{M}|^2 \right) \mathbf{q}  f_2 (q+m_2\mathbf{v}_2),
\end{align}
which is odd in $\mathbf{q}$ and so
\begin{align}
    \boxed{\bm{\mathcal{S}}_{\mathrm{decay},2} = 0}.
\end{align}
For the specific internal energy source term, we get 
\begin{align}
\tilde{\mathcal{S}}_{\mathrm{decay},2} &= - \frac{1}{\rho_2} \int d\Pi_p d\Pi_q \prod d\Pi_{k_j} (2 \pi)^4 \delta^4 \left(q - p - \sum_j k_j\right) \left( \sum |\mathcal{M}|^2 \right) \nonumber \\ 
& \qquad \qquad \times \left(\frac{|\mathbf{q}|^2}{2m_2^2} - e_2 \right)  m_2 f_2 (q + m_2 \mathbf{v}_2),
\end{align}
where the first term in the brackets is the definition of $e_2$, and so
\begin{align}
    \boxed{\tilde{\mathcal{S}}_{\mathrm{decay},2} = 0}.
\end{align}

\section{Upscattering collision terms} \label{app:collision_upscattering}

For the upscattering process $\chi_1 \chi_1 \to \chi_2\chi_2$, we start with
\begin{align}
    \mathcal{S}_{\mathrm{up}, 1} &= - \int d\Pi_{p} d\Pi_{p'} d\Pi_q d\Pi_{q'} (2 \pi)^4 \delta^4(p + p' - q - q') \left( \sum |\mathcal{M}|^2 \right) m_1 f_1(p) f_1(p'), \label{eq:dens_up_1_start} \\
    \bm{\mathcal{S}}_{\mathrm{up}, 1} &= - \frac{1}{\rho_1} \int d\Pi_{p} d\Pi_{p'} d\Pi_q d\Pi_{q'} (2 \pi)^4 \delta^4(p + p' - q - q') \left( \sum |\mathcal{M}|^2 \right) (\mathbf{p}-m_1 \mathbf{v}_1) f_1(p) f_1(p'), \label{eq:velocity_up_1_start} \\
    \tilde{\mathcal{S}}_{\mathrm{up}, 1} &= - \frac{1}{\rho_1} \int d\Pi_{p} d\Pi_{p'} d\Pi_q d\Pi_{q'} (2 \pi)^4 \delta^4(p + p' - q - q') \left( \sum |\mathcal{M}|^2 \right) \nonumber \\
    &\qquad \qquad \times \left(\frac{|\mathbf{p}-m_1 \mathbf{v}_1|^2}{2m_1^2} - e_1\right) m_1 f_1(p) f_1(p'), \label{eq:energy_up_1_start}
\end{align}
for the ground states, and 
\begin{align}
    \mathcal{S}_{\mathrm{up}, 2} &= + \int d\Pi_{p} d\Pi_{p'} d\Pi_q d\Pi_{q'} (2 \pi)^4 \delta^4(p + p' - q - q') \left( \sum |\mathcal{M}|^2 \right) m_2 f_1(p) f_1(p'), \label{eq:dens_up_2_start} \\
    \bm{\mathcal{S}}_{\mathrm{up}, 2} &= + \frac{1}{\rho_2} \int d\Pi_{p} d\Pi_{p'} d\Pi_q d\Pi_{q'} (2 \pi)^4 \delta^4(p + p' - q - q') \left( \sum |\mathcal{M}|^2 \right) (\mathbf{q}-m_2 \mathbf{v}_2) f_1(p) f_1(p'), \label{eq:velocity_up_2_start} \\
    \tilde{\mathcal{S}}_{\mathrm{up}, 2} &= + \frac{1}{\rho_2} \int d\Pi_{p} d\Pi_{p'} d\Pi_q d\Pi_{q'} (2 \pi)^4 \delta^4(p + p' - q - q') \left( \sum |\mathcal{M}|^2 \right) \nonumber \\
    &\qquad \qquad \times \left(\frac{|\mathbf{q}-m_2 \mathbf{v}_2|^2}{2m_2^2} - e_2\right) m_2 f_1(p) f_1(p'), \label{eq:energy_up_2_start}
\end{align}
for the excited states. 

\subsection{Upscattering loss terms for the ground state}

\subsubsection{Density}

Similar to before, we begin with the source term for number density in Eq. \eqref{eq:dens_up_1_start}, given by
\begin{align}
    \mathcal{S}_{\mathrm{up},1} &= - m_1 \int \frac{d^3 \mathbf{p}}{(2\pi)^3 2 E_p} \frac{d^3 \mathbf{p}'}{(2\pi)^3 2 E_{p'}} \frac{d^3 \mathbf{q}}{(2\pi)^3 2 E_q} \frac{d^3 \mathbf{q}'}{(2\pi)^3 2 E_{q'}} (2 \pi)^4 \delta^4(q + q' - p - p') \nonumber \\
    &\qquad \qquad \times \left( \sum |\mathcal{M}|^2 \right) f_1 (p) f_1(p').
\end{align}
We wish to relate the integrand of this quantity to the velocity-dependent inelastic scattering cross section $\sigma$. Note that
\begin{align}
    \int \frac{d^3\mathbf{q}}{(2\pi)^3 2E_q} \frac{d^3\mathbf{q}'}{(2\pi)^3 2E_q'} (2\pi)^4 \delta^4(q + q' - p - p') |\mathcal{M}|^2 = 4 E_p E_{p'} v_\mathrm{rel} \sigma(s), \label{eq:sigma_in_identity}
\end{align}
where $v_\mathrm{rel} = |\mathbf{u} - \mathbf{u}'|$ is the relative velocity\footnote{Note that this is valid only in the non-relativistic case. For relativistic scenarios, one must consider a more complicated expression for the M\o ller relative velocity, see $e.g.$ Ref.~\cite{Gondolo:1990dk}}, $\sigma$ is the inelastic scattering cross section, and $s = (p+p')^2 \approx 4m_1^2 (1+ v_\mathrm{rel}^2/4)$ is the usual Mandelstam variable in the non-relativistic limit, keeping only quadratic terms in velocity. Note that we assume the cross section $\sigma$ encapsulates any prefactors arising due to identical particles. Therefore we have that
\begin{align}
    \mathcal{S}_{\mathrm{up},1} &= - m_1 \int \frac{d^3 \mathbf{p}}{(2\pi)^3 } \frac{d^3 \mathbf{p}'}{(2\pi)^3 }  v_\mathrm{rel} \sigma(v_\mathrm{rel}) f_1 (p) f_1(p').
\end{align}
Next, converting from momentum-space to velocity-space, we have that the Maxwell-Boltzmann distributions can be converted by 
\begin{align}
    \int \frac{d^3 \mathbf{p}}{(2\pi)^3} f_1(\mathbf{p}) &= \int \frac{d^3 \mathbf{p}}{(2\pi)^3} \frac{n_1}{(2\pi m_1^2 \nu_1^2 )^{3/2}} \exp\left(-\frac{|\mathbf{p}|^2}{2 m_1^2 \nu_1^2}\right) \nonumber \\
    &= \int d^3 \mathbf{u}\frac{n_1}{(2\pi \nu_1^2 )^{3/2}} \exp\left(-\frac{|\mathbf{u}|^2}{2 \nu_1^2}\right) \nonumber \\
    &= \int d^3 \mathbf{u} ~f_1(\mathbf{u}) .
    \end{align}
In the non-relativistic limit writing $\mathbf{p} = m_1\mathbf{u}$ and $\mathbf{p}' = m_1 \mathbf{u}'$,
\begin{align}
    \mathcal{S}_{\mathrm{up},1} &= - m_1 \int d^3 \mathbf{u}~ d^3 \mathbf{u}' ~ v_\mathrm{rel} \sigma(v_\mathrm{rel}) \frac{n_1^2}{(2\pi)^{3} \nu_1^6} \exp\left(-\frac{|\mathbf{u}|^2 + |\mathbf{u}'|^2}{2 \nu_1^2}\right) .
\end{align}
We make the substitution $\mathbf{U} = (\mathbf{u} + \mathbf{u}')/2$ and $\mathbf{v}_\mathrm{rel} = \mathbf{u} - \mathbf{u}'$ so that
\begin{align}
    \mathcal{S}_{\mathrm{up},1} =-m_1 \int d^3 \mathbf{U}~ d^3 \mathbf{v}_\mathrm{rel}  ~v_\mathrm{rel}  \sigma(v_\mathrm{rel}) \frac{n_1^2}{(2\pi)^{3} \nu_1^6} \exp\left(-\frac{|\mathbf{U}|^2}{\nu_1^2} - \frac{v_\mathrm{rel}^2}{4\nu_1^2}\right).
\end{align}
Since $\sigma$ and $\vrel$ do not depend on $\mathbf{U}$, we can evaluate the integrals over $\mathbf{U}$ to get a factor of $\pi^{3/2} \nu_1^3$, so that
\begin{align}
    \mathcal{S}_{\mathrm{up},1} =- m_1\int d^3 \mathbf{v}_\mathrm{rel} ~  v_\mathrm{rel} \sigma(v_\mathrm{rel}) \frac{  n_1^2}{8 \pi^{3/2} \nu_1^3} \exp\left(- \frac{v_\mathrm{rel}^2}{4\nu_1^2}\right) = -\frac{m_1 n_1^2}{2 \sqrt{\pi} \nu_1^3} \int_0^\infty d v_\mathrm{rel}~ v_\mathrm{rel}^3 \sigma(v_\mathrm{rel})\exp\left(- \frac{v_\mathrm{rel}^2}{4\nu_1^2}\right).
\end{align}
Hence the loss term for the density evolution for $\chi_1$ in the presence of $\chi_1 \chi_1 \to \chi_2 \chi_2$ upscattering is given by
\begin{align}
    \boxed{\mathcal{S}_{\mathrm{up},1} = - \frac{\rho_1^2}{2 \sqrt{\pi} \nu_1^3} \int_0^\infty d v_\mathrm{rel}~ v_\mathrm{rel}^3 \frac{\sigma(v_\mathrm{rel})}{m_1} \exp\left(- \frac{v_\mathrm{rel}^2}{4\nu_1^2}\right)}.
\end{align}

\subsubsection{Velocity}

To find the evolution of the bulk velocity, we repeat a similar process as before but with $\mathbf{u} - \mathbf{v}_1$ in the integrand. Performing a shift once more so that $\mathbf{u} \to \mathbf{u} + \mathbf{v}_1$ and $\mathbf{u}' \to \mathbf{u}' + \mathbf{v}_1$, we have
\begin{align}
    \bm{\mathcal{S}}_{\mathrm{up}, 1} &= - \frac{m_1}{\rho_1} \int d^3 \mathbf{U}~ d^3 \mathbf{v}_\mathrm{rel}  ~v_\mathrm{rel}  \sigma(v_\mathrm{rel}) \frac{n_1^2}{(2\pi)^{3} \nu_1^6} (\mathbf{U} + \mathbf{v}_\mathrm{rel}/2) \exp\left(-\frac{|\mathbf{U}|^2}{\nu_1^2} - \frac{v_\mathrm{rel}^2}{4\nu_1^2}\right). \label{eq:chi1_vsource_upscatter}
\end{align}
The first term in the integrand vanishes since the integral over $\mathbf{U}$ is odd, whereas the second term in the integral vanishes because the integral is even in $v_\mathrm{rel}$ aside from the $\mathbf{v}_\mathrm{rel}$ factor, making the overall integral odd. Thus we have that
\begin{align}
    \boxed{\bm{\mathcal{S}}_{\mathrm{up}, 1} = 0}.
\end{align}

\subsubsection{Energy}

For the internal energy of the system, we must compute the higher moment, with $|\mathbf{u} - \mathbf{v}_1|^2/2 - e_1$ in the integrand. Again shifting the velocities, the dependence on $\mathbf{v}_1$ drops out, and we get
\begin{align}
    \tilde{\mathcal{S}}_{\mathrm{up}, 1} = -\frac{m_1}{\rho_1} \int d^3 \mathbf{U}~ d^3 \mathbf{v}_\mathrm{rel}  ~v_\mathrm{rel}  \sigma(v_\mathrm{rel}) \frac{n_1^2}{(2\pi)^{3} \nu_1^6} \left( \frac{|\mathbf{U} + \mathbf{v}_\mathrm{rel}/2 |^2}{2} - e_1 \right) \exp\left(-\frac{|\mathbf{U}|^2}{\nu_1^2} - \frac{v_\mathrm{rel}^2}{4\nu_1^2}\right).
\end{align}
Expanding $|\mathbf{U} + \mathbf{v}_\mathrm{rel}/2|^2 = |\mathbf{U}|^2 + |\mathbf{v}_\mathrm{rel}|^2/4 + \mathbf{U} \cdot \mathbf{v}_\mathrm{rel} $, we have that the term with the dot product goes to zero due to the integral being odd in $\mathbf{U}$. Thus we have
\begin{align}
    \tilde{\mathcal{S}}_{\mathrm{up}, 1} = - \int d^3 \mathbf{U}~ d^3 \mathbf{v}_\mathrm{rel}  ~v_\mathrm{rel}  \sigma(v_\mathrm{rel}) \frac{n_1}{(2\pi)^{3} \nu_1^6} \left( \frac{|\mathbf{U}|^2}{2} +   \frac{|\mathbf{v}_\mathrm{rel}|^2}{8} - e_1 \right) \exp\left(-\frac{|\mathbf{U}|^2}{\nu_1^2} - \frac{v_\mathrm{rel}^2}{4\nu_1^2}\right).
\end{align}
These Gaussian integrals over $\mathbf{U}$ can be easily integrated. Substituting in $e_1 = 3\nu_1^2/2$ and simplifying, we get
\begin{align}
    \boxed{ \tilde{\mathcal{S}}_{\mathrm{up}, 1} = - \frac{\rho_1}{16 \sqrt{\pi} \nu_1^3} \int_0^\infty dv_\mathrm{rel}  ~v_\mathrm{rel}^3  \frac{\sigma(v_\mathrm{rel})}{m_1}  \left( v_\mathrm{rel}^2 - 6 \nu_1^2  \right) \exp\left(- \frac{v_\mathrm{rel}^2}{4\nu_1^2}\right) } .
\end{align}

\subsection{Upscattering gain terms for the excited state}

\subsubsection{Density}

The gain term for the density evolution of $\chi_2$ is given by
\begin{align}
    \mathcal{S}_{\mathrm{up}, 2} &= \int \frac{d^3 \mathbf{p}}{(2\pi)^3 2 E_p} \frac{d^3 \mathbf{p}'}{(2\pi)^3 2 E_{p'}} \frac{d^3 \mathbf{q}}{(2\pi)^3 2 E_q} \frac{d^3 \mathbf{q}'}{(2\pi)^3 2 E_{q'}} (2 \pi)^4 \delta^4(q + q' - p - p') \left( \sum |\mathcal{M}|^2 \right) m_2 f_1 (p) f_1(p'),
\end{align}
where a similar procedure as before immediately gives 
\begin{align}
    \mathcal{S}_{\mathrm{up}, 2} = \frac{m_2 \rho_1^2}{2 \sqrt{\pi} m_1 \nu_1^3} \int_0^\infty d v_\mathrm{rel}~ v_\mathrm{rel}^3 \frac{\sigma(v_\mathrm{rel})}{m_1}\exp\left(- \frac{v_\mathrm{rel}^2}{4\nu_1^2}\right).
\end{align}
Taking the limit where $m_1 \approx m_2$, we have
\begin{align}
    \boxed{\mathcal{S}_{\mathrm{up}, 2} = \frac{ \rho_1^2}{2 \sqrt{\pi}  \nu_1^3} \int_0^\infty d v_\mathrm{rel}~ v_\mathrm{rel}^3 \frac{\sigma(v_\mathrm{rel})}{m_1}\exp\left(- \frac{v_\mathrm{rel}^2}{4\nu_1^2}\right)}.
\end{align}

\subsubsection{Velocity}

Unlike the velocity source term due to upscattering for $\chi_1$, for the velocity source term in \cref{eq:velocity_up_2_start} for $\chi_2$, the moment is measured with respect to the excited state velocity $\mathbf{v}_2$ while the Maxwellian distribution is centered around $\mathbf{v}_1$. Therefore we cannot simply shift by $\mathbf{v}_2$ to make the source term vanish. For convenience, we define $d \mathcal{W}$ such that $\mathcal{S}_{\rm up, 2} = m_2 \int d\mathcal{W}$. Explicitly, 
\begin{align}
    d\mathcal{W} \equiv \frac{d^3 \mathbf{p}}{(2\pi)^3 2 E_p} \frac{d^3 \mathbf{p}'}{(2\pi)^3 2 E_{p'}} \frac{d^3 \mathbf{q}}{(2\pi)^3 2 E_q} \frac{d^3 \mathbf{q}'}{(2\pi)^3 2 E_{q'}} (2 \pi)^4 \delta^4(q + q' - p - p') \left( \sum |\mathcal{M}|^2 \right)  f_1 (p) f_1(p'). \label{eq:dW_explicit}
\end{align}
Rewriting the velocity source term in \cref{eq:velocity_up_2_start} in terms of $d \mathcal{W}$,
\begin{align}
    \bm{\mathcal{S}}_{\mathrm{up,2}} = \frac{1}{\rho_2} \int d \mathcal{W} ~ (\mathbf{q} - m_2 \mathbf{v}_2).
\end{align}
Then, we remove the dependence on $\mathbf{q}$ by using momentum conservation $\mathbf{p} + \mathbf{p}' = \mathbf{q} + \mathbf{q}'$ and the fact that the collision integral is symmetric under $\mathbf{q} \leftrightarrow \mathbf{q}'$ and $\mathbf{p} \leftrightarrow \mathbf{p}'$. Therefore
\begin{align}
    \int d \mathcal{W} ~\mathbf{q} = \frac{1}{2} \int d\mathcal{W} ~ (\mathbf{q} + \mathbf{q}') = \frac{1}{2} \int d\mathcal{W} ~ (\mathbf{p} + \mathbf{p}') = \int d \mathcal{W} ~ \mathbf{p} = \int d \mathcal{W} ~ m_1 \mathbf{u}, \label{eq:momentum_identity}
\end{align}
Similar to before, since the Maxwell-Boltzmann distributions are centered around $\mathbf{v}_1$, we shift by $\mathbf{u} \to \mathbf{u} + \mathbf{v}_1$ and $\mathbf{u}' \to \mathbf{u}' + \mathbf{v}_1$. Therefore
\begin{align}
    \int d\mathcal{W} ~ \mathbf{q} = \int d \mathcal{W} ~ m_1 \left( \mathbf{U} + \mathbf{v}_1 + \frac{\mathbf{v}_{\rm rel}}{2} \right) ,
\end{align}
where we used the definitions $\mathbf{U} = (\mathbf{u} + \mathbf{u}')/2$ and $\mathbf{v}_{\rm rel} = \mathbf{u} - \mathbf{u}'$ as before. The shifted integrals over $\mathbf{U}$ and $\mathbf{v}_{\rm rel}$ vanish due to the odd parity of the integral (cf.~\cref{eq:chi1_vsource_upscatter}), and so we write the velocity source term for $\chi_2$ as 
\begin{align}
    \rho_2 \bm{\mathcal{S}}_{\mathrm{up,2}} = \int d \mathcal{W} ~ ( m_1 \mathbf{v}_1 - m_2 \mathbf{v}_2) = (m_1 \mathbf{v}_1 - m_2 \mathbf{v}_2) \int d \mathcal{W} = \frac{1}{m_2} (m_1 \mathbf{v}_1 - m_2 \mathbf{v}_2) \mathcal{S}_{\rm up, 2} .
\end{align}
In the limit where $m_1 \approx m_2$ once again, we then arrive at
\begin{align}
    \boxed{ \bm{\mathcal{S}}_{\mathrm{up,2}} =  \frac{ \rho_1^2 \mathbf{v}_{12} }{2 \rho_2 \sqrt{\pi}  \nu_1^3} \int_0^\infty d v_\mathrm{rel}~ v_\mathrm{rel}^3 \frac{\sigma(v_\mathrm{rel})}{m_1}\exp\left(- \frac{v_\mathrm{rel}^2}{4\nu_1^2}\right) }.
\end{align}

\subsubsection{Energy}
For the evolution of the internal energy of $\chi_2$, we consider the higher moment $|\mathbf{q} - m_2 \mathbf{v}_2|^2/{2m_2^2} - e_2$. Using the definition of $d\mathcal{W}$ in \cref{eq:dW_explicit}, the source term can be written compactly as 
\begin{align}
    \tilde{\mathcal{S}}_{\mathrm{up}, 2} = \frac{1}{\rho_2} \int d \mathcal{W}  \left(\frac{|\mathbf{q}-m_2 \mathbf{v}_2|^2}{2m_2} - m_2 e_2\right) . \label{eq:source_upscattering_2_compact}
\end{align}
In order to use the relationship in Eq. \eqref{eq:sigma_in_identity} we must eliminate the dependence on $\mathbf{q}$ in the non-relativistic limit. Note that $E_p + E_{p'} = E_q + E_{q'}$, and that $E_{p} \approx m_1 + |\mathbf{p}|^2/(2m_1)$ for each of the energies. Therefore integrating over the conservation of energy,
\begin{align}
    \int d\mathcal{W} \left[ \frac{|\mathbf{q}|^2}{2m_2} + \frac{|\mathbf{q}'|^2}{2m_2} \right] = \int d \mathcal{W} \left[ \frac{|\mathbf{p}|^2}{2m_1} + \frac{|\mathbf{p}'|^2}{2m_1} - 2 \delta \right], 
\end{align}
where again $\delta = m_2 - m_1$. Since the incoming (and outgoing) particles are identical, the momenta under the integrals on each side can be collected, so that
\begin{align}
    \int d \mathcal{W} ~ \frac{|\mathbf{q}|^2}{2m_2} = \int d\mathcal{W}  \left[\frac{|\mathbf{p}|^2}{2m_1} - \delta \right] . \label{eq:energy_identity}
\end{align}
Using this and expanding out the first term in \cref{eq:source_upscattering_2_compact}, we see that
\begin{align}
    \int d \mathcal{W}~ \frac{|\mathbf{q} - m_2 \mathbf{v}_2|^2}{2 m_2} &= \int d \mathcal{W} ~ \frac{\left[|\mathbf{q}|^2  + m_2^2 |\mathbf{v}_2|^2 - 2 m_2 \mathbf{q} \cdot \mathbf{v}_2 \right]}{2m_2} \nonumber \\
    &= \int d \mathcal{W} ~ \left[\frac{|\mathbf{p}|^2}{2m_1} - \delta + \frac{m_2|\mathbf{v}_2|^2}{2} - \mathbf{p} \cdot \mathbf{v}_2 \right] ,
\end{align}
where we used \cref{eq:momentum_identity} and \cref{eq:energy_identity}. Simplifying further by completing the square, we get
\begin{align}
    \tilde{\mathcal{S}}_{\mathrm{up}, 2} = \frac{1}{\rho_2} \int d \mathcal{W} ~ \left[\frac{|\mathbf{p} - m_1 \mathbf{v}_2|^2}{2m_1}  - \delta + \frac{\delta |\mathbf{v}_2|^2}{2} - m_2 e_2 \right] .
\end{align}
Explicitly, this is written as 
\begin{align}
    \tilde{\mathcal{S}}_{\mathrm{up}, 2} &= \frac{1}{\rho_2} \int \frac{d^3 \mathbf{p}}{(2\pi)^3 2 E_p} \frac{d^3 \mathbf{p}'}{(2\pi)^3 2 E_{p'}} \frac{d^3 \mathbf{q}}{(2\pi)^3 2 E_q} \frac{d^3 \mathbf{q}'}{(2\pi)^3 2 E_{q'}} (2 \pi)^4 \delta^4(q + q' - p - p') \nonumber \\
    &\qquad \qquad \times \left( \sum |\mathcal{M}|^2 \right) f_1 (p) f_1(p') \left(\frac{|\mathbf{p} - m_1 \mathbf{v}_2|^2}{2 m_1} - \delta + \frac{\delta |\mathbf{v}_2|^2}{2} - m_2e_2 \right). \label{eq:S_up_2_midway}
\end{align}
In the following, we omit the $\delta |\mathbf{v}_2|^2/2$ term as it is a higher order velocity correction, since for non-relativistic upscattering near threshold, $v \sim \sqrt{\delta/m_1}$, and so the entire term is of the order $m_1 v^4$.
Like in the previous cases, we shift all velocities by $\mathbf{v}_1$, and so again with $\mathbf{p} = m_1 \mathbf{u}$, the first term in the integrand is shifted as  $|\mathbf{p} - m_1 \mathbf{v}_2|^2 \to m_1^2 |\mathbf{u} +  \mathbf{v}_{12}|^2$. Therefore we simplify as
\begin{align}
    \tilde{\mathcal{S}}_{\mathrm{up}, 2} &= \frac{m_2}{\rho_2} \int d^3 \mathbf{u}~ d^3 \mathbf{u}' ~ v_\mathrm{rel} \sigma(v_\mathrm{rel}) \frac{n_1^2}{(2\pi)^{3} \nu_1^6} \left( \frac{ m_1}{2 m_2} |\mathbf{u} + \mathbf{v}_{12}|^2 - \frac{\delta}{m_2} - e_2 \right) \exp\left(-\frac{|\mathbf{u} |^2 + |\mathbf{u}'|^2}{2 \nu_1^2}\right),
\end{align}
Expanding out the $|\mathbf{u} + \mathbf{v}_{12}|^2$, we see that only the squared terms survive, since any term linear in $\mathbf{u}$ vanishes due to symmetry as before. Therefore after a change in coordinates similar to before, we arrive at
\begin{align}
    \tilde{\mathcal{S}}_{\mathrm{up}, 2} &= \frac{m_2}{\rho_2} \int d^3 \mathbf{U}~ d^3 \mathbf{v}_\mathrm{rel}  ~v_\mathrm{rel}  \sigma(v_\mathrm{rel}) \frac{n_1^2}{(2\pi)^{3} \nu_1^6} \left( \frac{m_1}{m_2} \left( \frac{|\mathbf{U}|^2}{2} +   \frac{|\mathbf{v}_\mathrm{rel}|^2}{8} +  \frac{| \mathbf{v}_{12}|^2}{2} \right) - \frac{\delta}{m_2} - e_2 \right) \nonumber \\
    &\qquad \qquad  \times \exp\left(-\frac{|\mathbf{U}|^2}{\nu_1^2} - \frac{v_\mathrm{rel}^2}{4\nu_1^2}\right).
\end{align}
Evaluating the integral over $\mathbf{U}$,
\begin{align}
    \tilde{\mathcal{S}}_{\mathrm{up}, 2} = \frac{m_2 n_1^2}{8\pi^{3/2} \nu_1^3 \rho_2} \int  d^3 \mathbf{v}_\mathrm{rel}  ~ v_\mathrm{rel} \sigma(v_\mathrm{rel})   \left( \frac{m_1}{m_2} \left( \frac{3\nu_1^2}{4} +   \frac{|\mathbf{v}_\mathrm{rel}|^2}{8} +  \frac{| \mathbf{v}_{12}|^2}{2} \right) - \frac{\delta}{m_2} - e_2 \right) \exp\left(- \frac{v_\mathrm{rel}^2}{4\nu_1^2}\right),
\end{align}
which, when changing to spherical coordinates and rearranging, yields
\begin{align}
    \tilde{\mathcal{S}}_{\mathrm{up}, 2} = \frac{m_2 \rho_1^2}{16 \sqrt{\pi} \nu_1^3 m_1 \rho_2} \int_0^\infty  d v_\mathrm{rel}  ~v_\mathrm{rel}^3  \frac{\sigma(v_\mathrm{rel})}{m_1} \left( \frac{m_1}{m_2} \left( 6 \nu_1^2 +   v_\mathrm{rel}^2 +  4| \mathbf{v}_{12}|^2 \right) - 8 \frac{\delta}{m_2} - 8 e_2 \right) \exp\left(- \frac{v_\mathrm{rel}^2}{4\nu_1^2}\right).
\end{align}
Taking the limit for any fractions between $m_1$ and $m_2$ to be unity, we have
\begin{align}
    \boxed{\tilde{\mathcal{S}}_{\mathrm{up}, 2} = \frac{\rho_1^2}{16 \sqrt{\pi} \nu_1^3 \rho_2} \int_0^\infty  d v_\mathrm{rel}  ~v_\mathrm{rel}^3  \frac{\sigma(v_\mathrm{rel})}{m_1} \left(  6 \nu_1^2 +   v_\mathrm{rel}^2 +  4| \mathbf{v}_{12}|^2 - 8 \frac{\delta}{m_2} - 8 e_2 \right) \exp\left(- \frac{v_\mathrm{rel}^2}{4\nu_1^2}\right)}.
\end{align}
When writing \cref{eq:euler3_2} in the main text, we have used $m_2 \approx m_1$ to express the suite of fluid equations in terms of only $m_1$.

\section{Collapse time model fit}

\label{sec:collapse_model_fit}

Our ansatz for the collapse time for dSIDM halos relies on parameterizing the $\xi$-dependent departure from the universal evolution trajectory for SIDM halos.
In total, there are seven free parameters, listed in Table~\ref{tab:collapse_model_parameters}, collectively denoted by $\boldsymbol{\theta}$.

To fix these free parameters in our model, we perform a fit to the ensemble of 200 dissipative simulations described in Sec.~\ref{sec:dissipative_collapse_timescale}.
From each simulation, we extract the actual core density $\rho_0^{(i)}$ and velocity dispersion $\nu_0^{(i)}$ (at the maximal core time), and collapse time $t_{\rm coll}^{(i)}$ for each simulation $i$.
Since the elastic-only collapse time $t_{\rm el}$ takes a fixed common value for all halos, we also determine $\xi^{(i)} = t_{\rm coll}^{(i)}/t_{\rm el}$ for all simulated halos.
Next, for each $\boldsymbol{\theta}$, we predict these quantities according to our model, for each simulation.
These are denoted by hatted quantities $\hat{\rho}_0^{(i)}$, $\hat{\nu}_0^{(i)}$, and $\hat{t}_{\rm coll}^{(i)}$.

We define a loss function $\mathcal{L}$ as follows:
\be
\mathcal{L}(\boldsymbol{\theta}) 
= 
\sum_i \left[ w_t^{(i)} \log^2\left({t}_{\rm coll}^{(i)}/\hat{t}_{\rm coll}^{(i)}\right) 
+ w_\rho \log^2\left({\rho}_{0}^{(i)}/\hat{\rho}_{0}^{(i)}\right)
+ w_\nu \log^2\left({\nu}_{0}^{(i)}/\hat{\nu}_{0}^{(i)}\right) \right] \, ,
\ee
where the sum runs over all simulations.
Naturally, the loss term vanishes where the simulations and model agree and becomes large where the measured and predicted quantities diverge.
The presence of logarithms is motivated by the fact that these quantities can differ by orders of magnitude across different simulations.
We include the following weight factors
\be
w_t^{(i)} = (1 - \hat{\xi}^{(i)})^2 \, , \quad
w_\rho = w_\nu = 0.1 \, ,
\ee
so that our fit is mostly driven to fit $t_{\rm coll}$ in strongly cooled halos, while weakly cooled halos are fit trivially in our model anyway. 
Our results are not strongly dependent on $w_\rho$ and $w_\nu$ and the values chosen suffice for the purposes of this paper.
The best fit values reported in Table~\ref{tab:collapse_model_parameters} are obtained by minimizing $\mathcal{L}$.

\end{document}